\documentclass[preprint2]{aastex}
\def\stacksymbols #1#2#3#4{\def\theguybelow{#2}
        \def\verticalposition{\lower#3pt}
        \def\spacingwithinsymbol{\baselineskip0pt\lineskip#4pt}
        \mathrel{\mathpalette\intermediary#1}}
\def\intermediary #1#2{\verticalposition\vbox{\spacingwithinsymbol
        \everycr={}\tabskip0pt
        \halign{$\mathsurround0pt#1\hfil##\hfil$\crcr#2\crcr
                \theguybelow\crcr}}}
\def\lta{\stacksymbols{<}{\sim}{2.5}{.2}}
\def\gta{\stacksymbols{>}{\sim}{3}{.5}}

\shorttitle{Infrared Properties of E and S0 Galaxies}
\shortauthors{Temi, Brighenti and Mathews}

\begin{document}


\title{SPITZER OBSERVATIONS OF PASSIVE AND STAR FORMING
EARLY-TYPE GALAXIES:\\
AN INFRARED COLOR-COLOR SEQUENCE}

\author{Pasquale Temi\altaffilmark{1,2},
Fabrizio Brighenti\altaffilmark{3,4}, William
G. Mathews\altaffilmark{3} }
\altaffiltext{1}{Astrophysics Branch, NASA/Ames Research Center, MS
  245-6,
Moffett Field, CA 94035 (pasquale.temi@nasa.gov).}
\altaffiltext{2}{
Department of Physics and Astronomy, University of Western
Ontario,
London, ON N6A 3K7, Canada. }
\altaffiltext{3}{University of California Observatories/Lick
  Observatory,
Board of Studies in Astronomy and Astrophysics,
University of California, Santa Cruz, CA 95064
(mathews@ucolick.org).}
\altaffiltext{4}{Dipartimento di Astronomia,
Universit\`a di Bologna, via Ranzani 1, Bologna 40127, Italy
(fabrizio.brighenti@unibo.it).}

\begin{abstract}
We describe the infrared properties of a large sample of early type 
galaxies, comparing data from the Spitzer archive with 
$Ks$-band emission from 2MASS.
While most representations of this data result in 
correlations with large scatter, we find a remarkably tight 
relation among colors formed by ratios of luminosities in Spitzer-MIPS
bands (24, 70 and 160$\mu$m) and the $Ks$-band.
Remarkably, this correlation among E and S0 galaxies 
follows that of nearby normal galaxies
of all morphological types.
In particular, the tight infrared color-color correlation for S0 galaxies 
alone follows that of the entire Hubble sequence of normal galaxies,
roughly in order of galaxy type from ellipticals to spirals to irregulars.
The specific star formation rate of S0 galaxies 
estimated from the 24$\mu$m luminosity 
increases with decreasing $K$-band luminosity (or stellar mass) 
from essentially zero, as with most massive ellipticals, 
to rates typical of irregular galaxies.
Moreover, the luminosities of the many infrared-luminous S0 galaxies 
can significantly exceed
those of the most luminous (presumably post-merger) E galaxies.
Star formation rates in the most infrared-luminous S0 galaxies 
approach 1--10 solar masses per year.
Consistently with this picture we find that while most early-type galaxies
populate an infrared red sequence, about 24\% of the objects (mostly S0s)
are in an infrared blue cloud together with late type galaxies.
For those early-type galaxies also observed at radio frequencies 
we find that the far-infrared luminosities correlate 
with the mass of neutral and molecular hydrogen, 
but the scatter is large.
This scatter suggests that the star formation may be intermittent 
or that similar S0 galaxies with 
cold gaseous disks of nearly equal mass 
can have varying radial column density distributions  
that alter the local and global star formation rates.
\end{abstract}

\vskip.1in

\section{Introduction}

If S0 galaxies did not exist it is possible that they would 
not have been missed. 
Late type, star-forming galaxies including spirals are expected in 
early galactic evolution and elliptical galaxies are thought to 
form by mergers among these ancient late type galaxies or, at a later stage,  
from dry mergers with each other. 
S0 or lenticular galaxies are
differentiated from these more predictable and recognizable galaxies 
by being generally structureless 
with stellar bulges and rotating disks having light distributions that 
are somewhat less concentrated than in ellipticals.
The distinction between E and S0 morphologies is not always 
easy, particularly when S0s are viewed face-on.
This classification difficulty has resulted in the
notion of E-S0 galaxies where the uncertain type may 
either be an intrinsic reality or a consequence of 
the limitations of the astronomical classifier.
S0s are often a nuisance in large surveys since 
the S0-E distinction decreases with redshift and 
the two types of galaxies are often combined.
  
However, many local S0 galaxies are quite distinct 
from ellipticals, particularly as a result of their significant 
gas content, star formation rates, and optical colors.
At infrared wavelengths many S0 galaxies exhibit 
a range of unusual properties 
not often found in ellipticals.

Recently we discussed the infrared properties of 
34 S0 and E galaxies 
in the SAURON sample that have also been observed in the 
infrared with the Spitzer telescope 
(Temi, Brighenti \& Mathews 2009; hereafter TBM09). 
This relatively small Spitzer-SAURON sample is useful 
since SAURON galaxies have 
been intensively observed at many other wavelengths.
While the infrared emission from some S0 galaxies in the 
Spitzer-SAURON sample closely resembles gas-free elliptical 
galaxies, many S0 galaxies contain rotationally supported 
kpc-sized disks of star-forming cold gas 
(e.g. Combes, Young \& Bureau 2007;
Young. Bendo \& Lucero 2009) and dust 
that emit more strongly in Spitzer bandpasses than ellipticals. 
The star formation rates in S0s in the Spitzer-SAURON sample,
derived from their 24$\mu$m emission, 
are small, $\lta 0.2$ $M_{\odot}$ yr$^{-1}$, but correlate 
with the magnitude of the H$\beta$ spectral index 
produced by relatively young A-F stars (TBM09).
However, in SAURON S0 galaxies 
we found that star formation rates estimated 
from Balmer emission line luminosities 
are often much less than 
star formation rates found from 24$\mu$m emission.
The insufficient emission line evidence for OB stars 
is curious since slightly less massive young stars are 
often visible. 
In any case, hydrogen line emission is easily absorbed by dust,  
so estimates of star formation rates in S0 galaxies 
are probably more reliably determined from 24$\mu$m dust emission
or by a combination of $H\alpha$ and infrared luminosities
(Kennicutt et al. 2009)

In this paper we discuss infrared Spitzer observations 
of a much larger, more comprehensive sample of E and S0 
galaxies taken from the Spitzer archive.
This extended sample reveals some remarkable attributes of 
S0 galaxies. 
Most important of these is a very narrow correlation between
colors formed by ratios of mid and far-infrared 
Spitzer luminosities with $K$-band luminosities 
in which the dust properties differentiate many but not all S0s
from gas-free E galaxies.
Furthermore, it is remarkable that the color-color correlation
for S0s is nearly identical to that of normal galaxies spanning all
morphological types.
Although ancillary data is less abundant for this extended 
sample of early-type galaxies than for the SAURON-Spitzer sample, 
the mass of neutral and 
molecular hydrogen is known for many of them and this 
provides several interesting comparisons with  
star formation rates based on their Spitzer 24$\mu$m luminosities.

\section{Spitzer Observations of Early Type Galaxies}

Table 1 contains relevant information about our extended sample of
elliptical and lenticular galaxies found from the Spitzer archives. 
We searched the archives to select early-type galaxies 
within a distance of 30 Mpc that have been observed with
Spitzer at far infrared wavelengths.
About 60 more distant galaxies have been added
from known Spitzer observing programs and by querying large lists of
early-type galaxies known from other surveys.
Of the 225 galaxies in Table 1, 121 are ellipticals,
26 are E-S0 and 76 are S0. 
Two galaxies have uncertain classification
(NGC 3656 and NGC 5666). 
NGC 5666 was classified as an elliptical in early studies by Nilson (1973) and Lake et al. (1987). Recently Donzelli \& Davoust (2003) found evidence of spiral structures from deep CCD images, suggesting a late type Sc galaxy. However photometric analysis indicates a number of contradictory morphological properties that point toward a classification of an early type disk galaxy which has accreted a gas rich dwarf galaxy during a minor merger event.  
NGC 3656 is classified as (R')I0: pec by NED. Its morphological classification is not reported in the HyperLeda catalog, and is listed as non-Magellanic irregular in RC2 catalog. From near infrared photometry, Wiklind, Combes \& Henkel (1995) classify NGC3656 as a disturbed elliptical with an envelope of faint shells and a dust lane. They point out that it could be also considered a face-on S0.
None of our results changes if these two galaxy are removed from the sample.
This table also includes all early-type galaxies in our
proprietary Spitzer observations
(Temi, Brighenti \& Mathews 2007, hereafter TBM07)
as well as early-type SAURON galaxies observed with Spitzer
and discussed in TBM09.
All but four galaxies in Table 1 
are massive or relatively massive, i.e. ${\rm Log} L_{Ks} > 8.4$ 
($M_{Ks}<-17.6$).
Table 1 contains flux densities and specific luminosities
$L_{\lambda}$ at 24, 70 and 160$\mu$m.\footnotemark[1]

\footnotetext[1]{
In this paper $L_{\lambda}$ represents
$\lambda L_{\lambda}$ in erg s$^{-1}$.
This differs from our previous notation in which $L_{\lambda}$
represented the Spitzer band width multiplied by the specific
luminosity.}

\section{Observations and Reduction}

The FIR data presented here were obtained with the Multiband
Imaging Photometer (MIPS; Rieke et al. 2004) on board the
Spitzer Space Telescope (SST ; Werner et al. 2004) in three
wavebands centered at 24, 70, and 160$\mu$m. These data, collected
from the Spitzer public archive, do not form a homogeneous
and uniform data set in terms of image depth and observing
mode. 
Column 2 of Table 1 lists the program identification (ID) and principal
investigator (PI) for the various original Spitzer observing programs 
from which our sample galaxies are selected. The reader is
referred to these programs to obtain details on the observing
modes and imaging strategy, as well as the on-source integration
time for each target in the sample. As an example, Virgo Cluster
galaxies recorded under the guaranteed time program (PID 69,
PI G. Fazio) reach a relatively low sensitivity of 0.5 MJy sr$^{-1}$
and 1.1 MJy sr${-1}$ (1$\sigma$) at 70 and 160$\mu$m, respectively, while
other galaxies (i.e., PID 20171, PI P. Temi) have deeper maps
at a sensitivity level of only 0.12 MJy sr$-1$ and 0.3 MJy sr$-^{1}$ for
the same two wavebands. The Spitzer Infrared Nearby Galaxies
Survey (SINGS) data are recorded in MIPS scan mode, covering
a very large sky area ($30\arcmin \times  10\arcmin$), 
incorporating two separated
passes at each source location. SINGS images correspond to
maps with intermediate sensitivity. Apart from the SINGS
observations, data have been acquired in MIPS photometry
mode, allowing appropriate coverage of the sources and their
extended emission.

Reduction of data from the Spitzer archives follows the same procedure
used in Temi et al. (2007). Here, we briefly summarize
the basic processes involved. We started with
the Basic Calibrated Data (BCD) products from the Spitzer
Science pipeline (version 16.1) to construct mosaic images for
all objects. Final calibrated images have been produced using
the Mosaicking and Point-source Extraction (MOPEX) package
developed at the Spitzer Science Center (Makovoz et al.
2006). MOPEX includes all the functions and steps necessary
to process BCD data into corrected images and co-add them
into a mosaic. The major MOPEX pipeline used was the mosaic
pipeline which consists of a number of individual modules
to be run in sequence to properly perform the reduction. We
refer the reader to the MOPEX web page for a detailed description
 of each module. Since the data sets presented here have
been acquired in different observing modes, the modules chosen
to buildup the reduction flow and their parameter setup have
been carefully selected to properly remove mode-dependent
artifacts in the final mosaic (outlier detection and median
filtering). Data reduction presented here make use of the 
most updated MIPS flux calibration factors based on a 
large sample of stars and asteroids (Engelbracht et al. 2007;
Gordon et al. 2007; Stansberry et al 2007).

Foreground stars and background galaxies present
in the original mosaiced images were deleted before flux
extraction was performed. These were identified by eye and
cross-checked using surveys at other wavelengths (Digital
Sky Survey and 2MASS). Flux densities were extracted from
apertures that cover the entire optical disk (R25). Sky subtraction
was performed by averaging values from multiple apertures
placed around the target, avoiding any overlap with the faint extended
emission from the galaxy. Statistical uncertainties
related to sky subtraction are usually less than 1\% but can be
appreciable (tens of percent) for faint sources. Observed infrared
flux densities for each galaxy are listed in 
Table 1. Columns 10-12 
contain the specific flux $F_{\lambda}$ (mJy) identified by  
each MIPS wavelength 24, 70, and 160$\mu$m.

Systematics in the MIPS calibration result in fluxes uncertain
at the 5\% level at 24$\mu$m and 10\% at 70 and 15\% at 160$\mu$m. 
The uncertainties listed in Table 1 include the systematic
uncertainties. Aperture corrections for extended sources were
applied to the fluxes as described in the Spitzer Observer's
Manual. The corresponding MIPS luminosities are 
$L_{\lambda} = \lambda F_{\lambda} \cdot 4 \pi D^2$,
where distances $D$ are from Table 1.

\section{Infrared Photometry of E and S0 Galaxies}

Figure 1 shows the luminosities in three Spitzer bandpasses 
plotted against the $Ks$-band luminosity $L_{Ks}$. 
The tight correlation among the elliptical galaxies (filled red  
circles) in the upper panel is expected from an old stellar population 
in which most of the warmer dust emitting at 24$\mu$m 
is in circumstellar outflows from old red giant stars. 
The correlation visible in this panel is approximately
\begin{equation}
{\rm~Log} L_{24} \approx {\rm~Log} L_{Ks} + 30.1
\end{equation}
(TBM09).
(Here and in the following $L_{24}$ and other MIPS luminosities 
are in erg s$^{-1}$ and $L_{Ks}$ are in solar units $L_{Ks,\odot}$.)
A circumstellar origin for $L_{24}$ in the old stellar populations 
in elliptical galaxies is based on the $r^{1/4}$ 
surface brightness profiles observed 
in both 24$\mu$m and $Ks$-band emission 
(Temi, Brighenti \& Mathews 2008, hereafter TBM08).
Whenever possible, the optical colors of 
early type galaxies in Figure 1 are designated with red or blue 
colors if $U-V$ exceeds 1.1 or not. 
Blue E and S0 galaxies 
are found only for ${\rm Log}L_{Ks} \lta 11$, 
consistent with the much larger sample of 
Kannappan et al. (2009).

Relatively weak emission at 70 and 160$\mu$m from
elliptical galaxies can be interpreted as emission
from cold interstellar dust heated by diffuse starlight
and thermal electrons in the hot interstellar gas (TMB07).
Somewhat stronger emission at 70 and 160$\mu$m 
(as well as 24$\mu$m) indicates either 
emission from dust that is 
buoyantly transported out from galactic cores by AGN feedback 
or, for the most infrared-luminous ellipticals, 
emission from large masses of dusty, star-forming gas
apparently acquired in mergers with gas-rich galaxies.
While a few S0 galaxies (open circles) occupy the same regions as 
the elliptical galaxies in Figure 1, 
the infrared luminosities of many 
S0 galaxies exceed those of E galaxies 
in all three Spitzer passbands. 

While most representations of the Spitzer data result in 
broad scatter plots such as Figure 1, 
much tighter correlations emerge when the infrared luminosities 
are normalized by $L_{Ks}$ 
as in Figure 2.
Figure 2 significantly extends Figure 5 in TBM09\footnotemark[2]
which does not continue above
${\rm Log} L_{24}/L_{Ks} \approx 31$. 
In Figure 2 and subsequent figures we plot only Spitzer 
detections, not upper limits.

\footnotetext[2]{Apart from a constant, 
$\Delta{\rm Log} (\lambda L_{\lambda})_{24}$ 
in Figure 5 of TBM09 is
essentially identical to ${\rm Log}L_{24}/L_{Ks}$
used in this paper.}

The banana-shaped 
correlation in Figure 2 is driven by 
variations in the luminosity $L_{Ks}$ of old stars 
and varying amounts of star-forming dusty gas. 
Most of the E galaxies and a few S0 galaxies 
with ${\rm Log} L_{70}/L_{Ks}$ 
and ${\rm Log} L_{160}/L_{Ks}$ less than $\sim31$
have 24-Ks colors ${\rm Log} L_{24}/L_{Ks} \approx 30.1$
characteristic of the spectral energy distribution (SED)
of an old stellar population including  
circumstellar emission at 24$\mu$m (TBM09).
We expect ${\rm Log}L_{24}/L_{Ks}$ to asymptotically 
approach $\sim$30.1 as ${\rm Log}L_{FIR}/L_{Ks}$ 
decreases in Figure 2. 
It is unclear why the approach toward this final SED value is 
more complete in the horizontal distribution of red galaxies 
in the upper panel of Figure 2 than in the lower panel 
which may have a small residual positive slope when 
${\rm Log} L_{160}/L_{Ks} \lta 31$.
Provided the infrared colors $L_{24}/L_{70}$ and $L_{160}/L_{24}$
are reasonably constant,
galaxies with smaller $L_{Ks}$ lie toward the upper right
in both panels of Figure 2.
Figure 3 is a second version of the data in Figure 2
that shows in detail how $L_{Ks}$ varies along the correlations. 
While $L_{Ks}$ decreases systematically toward 
the upper right in both panels of Figure 3, 
this trend is far from monotonic. 

In Figure 2 we label the morphological type of early galaxies as 
either E or S0 while in practice 
morphological designations along the Hubble sequence 
are not rigorously discrete. 
In the Introduction we mentioned the uncertain morphological 
transition between E and S0 galaxies and the intermediate 
E-S0 type. 
As can be seen in Column 3 of Table 1, the transition from 
S0 to later type Sa galaxies is also gradual with S0-a (S0a or S0/a)
galaxies bridging the transition. 
To explore the importance of these intermediate galaxy types, 
we examined plots similar to 
Figure 2 with early type galaxies binned by 
their specific HyperLeda morphology: E, E-S0, S0 and S0-a. 
For the relatively small numbers of galaxies in each subtype, 
we were unable to detect any difference between the 
distributions of S0 and S0-a galaxies in 
Figure 2, both types appear to be spread evenly along 
the entire color-color plot. 
The small number ($\sim10$) of E-S0 galaxies in Table 1 
appear to be distributed as expected, 
sharing properties of both red and blue (E and S0) points 
in Figure 2. 
Consequently, in our plots we simply use 
the de Vaucouleurs T = -3.0 to separate E from S0 and 
regard all HyperLeda S0-a galaxies in Table 1 as S0 galaxies.
Such distinctions are likely to be difficult or impossible 
for early type galaxies at high redshift.

To investigate further 
the origin of the tight correlations in Figures 2 and 3, 
we show in Figures 4a and 4b the independent variations of 
$L_{Ks}$ and $L_{24}$ with 
${\rm Log} L_{FIR}/L_{Ks}$, 
plotted along the horizontal axis 
exactly as in Figures 2 and 3. 
Figure 4a clearly reveals that 
the correlations in Figure 2 are largely a consequence of 
a pronounced decline in $L_{Ks}$.
$L_{Ks}$ can be considered a proxy for 
the mass of the old stellar population. 
But Figure 4b shows that
the correlation in Figure 2 is also influenced by
an increase in $L_{24}$.
It is most interesting that the considerable scatter in
Figures 4a and 4b is 
greatly reduced in Figure 2.
Evidently values of both $L_{24}$ and $L_{Ks}$ 
for each galaxy conspire to ensure 
the much smaller scatter in Figure 2. 

When ${\rm Log} L_{24}/L_{Ks}$ sufficiently exceeds 
the old stellar population SED value 30.1 in Figure 2, 
${\rm Log} L_{24}/L_{Ks}$ becomes a measure of 
the specific star formation rate per unit $L_{Ks}$.
In TBM09 we modified the expression for the 
star formation rate (SFR) based on 24$\mu$m emission from 
Calzetti et al. (2007) so that it is forced to 
become zero 
for old elliptical galaxies where circumstellar dust 
produces the 24$\mu$m emission, not star formation: 
\begin{displaymath}
{\rm SFR}(M_{\odot},{\rm~yr}^{-1})~~~~~~~~~~~~~~~~~~~~~~~~~~~~~~~~~~~~
\end{displaymath}
\begin{equation}
~~~~~~~= 1.24 \times 10^{-38} 
[\max(0.0, L_{24} - 10^{30.1}L_{Ks})]^{0.885}.
\end{equation}
This equation only applies when 
${\rm Log} L_{24}/L_{Ks}$ exceeds the range of values 
observed in old E galaxies, i.e. $\rm{Log}L_{24}/L_{Ks} \gta 30.5$
or 
\begin{displaymath}
{\rm SFR}(M_{\odot},{\rm~yr}^{-1}) 
\gta 0.044 (L_{24}/10^{11} L_{Ks,\odot})^{0.885}.
\end{displaymath}
For those (mostly S0) galaxies in Figure 2 
with ${\rm Log} L_{24}/L_{Ks} \gta 30.5$,
the specific SFR per unit $L_{Ks}$ 
increases with increasing 
${\rm Log} L_{24}/L_{Ks}$.
This is consistent with the bluer colors of 
S0 galaxies in the upper part of the correlation in Figure 2
and with the ``downsizing'' notion that most current star formation
is occurring in low-$L_{Ks}$ 
galaxies (Cowie et al. 1996). 
Using equation (2), the star formation rates for 
the S0 galaxies having the largest $L_{24}$ 
(top panel of Figure 1) is remarkably large:
$1.4 \lta {\rm SFR} \lta 11$ $M_{\odot}$ yr$^{-1}$ for 
$43 \lta {\rm Log}L_{24} \lta 44$ erg s$^{-1}$. 

\section{Comparing Spitzer Observations of Early Type Galaxies
with a More Morphologically Diverse Sample of Normal Galaxies}

We now compare 
the results of our archival Spitzer sample of E and S0 galaxies with 
similar Spitzer observations of 
nearby normal galaxies spanning all morphological types. 
For this we use the SIRTF Nearby Galaxy Survey
or SINGS sample (Kennicutt et al. 2003)
which has been extensively observed with Spitzer and other 
telescopes. 
Using data from Kennicutt et al. (2003)
and the SINGS website,
we plot in Figure 5 the SINGS Spitzer data in the same format as 
shown in the upper panels of Figures 2 and 4.
In Figure 5 the SINGS galaxies are  
binned by morphological type designated by the 
HyperLeda designation of the 
de Vaucouleurs T parameter (upper panel) and 
by $L_{Ks}$ (lower panel)\footnotemark[3]. 

\footnotetext[3]{The following ten galaxies are common 
to both the SINGS sample and ours (Table 1): NGC 584, 855, 1266,
1316, 1377, 1404, 3265, 4125, 4552 and 5866.
In Figure 5 and subsequent
figures that include galaxies from both samples, 
we plot these nine galaxies as members only of our sample.}

Figure 5 reveals that the morphological type 
of SINGS galaxies becomes progressively later (larger $T$) 
and the stellar luminosity $L_{Ks}$ becomes smaller with 
increasing Log$L_{24}/L_{Ks}$, 
a measure of the star formation rate per $L_{Ks}$.
Some of the morphological bins scatter along the correlation 
more than others and, as expected from Figure 2, 
this scatter is particularly large for SINGS S0 galaxies.
Nevertheless, at least for some morphological types, 
it is remarkable that an approximate value 
of the type $T$ can be inferred  
from $L_{Ks}$ plus a single infrared observation of either
24, 70 or 160$\mu$m. 
This single-color morphological identification is particularly 
effective for E ($-5  < {\rm T} < -3$) 
and Sb-Sd ($3 < {\rm T} < 7$) galaxies 
where the variation of $L_{24}/L_{Ks}$ is more concentrated 
in Figure 5. 

Dale et al. (2007) 
describe the detailed spectral energy distributions of 
the SINGS galaxies from UV to sub-millimeter wavelengths. 
We note that our Figure 5 is a somewhat simplified version of
their Figure 10 where they plot a photometric measure of the 
specific star formation rate -- a combination of UV emission from 
young stars with reprocessed infrared emission normalized 
by $L_{Ks}$ --  
against the ratio of IR to UV light,
also identifying each galaxy by its morphological type.
Their figure lacks the banana simplicity of our Figure 5 
in part because old population stars in early type galaxies
can emit UV (and 24$\mu$m) emission that is 
unrelated to the specific star formation rate.

In Figure 6 we combine the color-color plot 
of our sample of early type galaxies (as in Figure 2) 
with the SINGS galaxies now represented simply with black $\times$ 
symbols. 
It is seen that galaxies of all morphological types and 
stellar masses lie along the same well defined 
banana-shaped correlation. 
In particular the S0 galaxies in our sample obey exactly the 
same correlation in the infrared color-color plot as the 
entire morphological sequence of normal galaxies, from ellipticals to 
irregulars. 
The coincidence of S0 infrared colors in Figure 6 with those 
of late-type SINGS galaxies, from spirals to irregulars, 
seems particularly surprising 
since cold gas in S0 galaxies is typically confined to 
small kpc-sized disks 
(e.g. Young, Bendo \& Lucero 2009).
Furthermore, S0 galaxies of varying $L_{Ks}$ have 
star formation rates similar to spiral or irregular galaxies that 
are nearby in the infrared color-color plot (Fig. 6). 
However, the specific SFR that we refer to is based on 
observations of ${\rm Log} L_{24}/L_{Ks}$ (eqn. 2), 
not Balmer emission line luminosities (Kennicutt et al. 2003) which, 
as least for the SAURON galaxies (TBM09), may not be as reliable
in estimating the SFRs in S0 galaxies as in 
normal late type galaxies (Kennicutt et al. 2003).

In the upper panel of 
Figure 7 we compare galaxies from our early-type and 
the SINGS samples 
in a different way by over-plotting both samples 
as in the upper panel of Figure 1 but with $L_{24}$
normalized with $L_{Ks}$.
Using the symbol notation of Figure 1, 
the sample of early-type galaxies is shown 
as filled (E and E-S0) or open (S0 and later types) circles which are red or blue 
depending on the $U-V$ color. 
Early-type galaxies with unknown $U-V$ are shown with green symbols.
The infrared color-magnitude diagram in the upper panel shows that 
galaxies are separated
in an infrared red sequence (the horizontal strip at the bottom,
locus of most ellipticals)
and an infrared blue cloud, where almost all the SINGS galaxies and many S0s reside.
The two sequences merge at the high luminosity end of our sample.
About 24\% of our early-type galaxies inhabit the infrared blue cloud.
This result is consistent with the fraction of {\it not passive} early-type galaxies
in the COMA cluster estimated by Clemens et al. (2009) through the analysis of
the ($K_s - [16\;\mu{\rm m}]$) color-magnitude diagram.

The SINGS galaxies (black $\times$ symbols in Fig. 7), 
occupy most of the same region as the early-type sample but 
extend to lower values of $L_{Ks}$, possibly reflecting a relative
absence of low mass S0 galaxies.
While this plot undoubtedly suffers from incompleteness and 
the multifarious nature of archived data, 
we see a gap or absence of galaxies with 
$30.5 \lta {\rm Log} L_{24}/L_{Ks} \lta 31.25$ in the range 
$8.5 \lta {\rm Log} L_{Ks} \lta 10$.
However, the star formation rate would be very small 
for any galaxy that might exist in this gap.
For example, at the  upper boundary of this region 
devoid of galaxies $L_{24} \approx 10^{31.25} L_{Ks}$ 
and by equation (2) this corresponds to 
$SFR < 0.005 (L_{Ks}/10^9 L_{Ks,\odot})^{0.885}$ 
$M_{\odot}$ yr$^{-1}$ 
which is very small indeed.

Figure 8 shows an optical color-magnitude diagram for our sample 
(and the SINGS sample) based on  
photometry from the website maintained by the 
Sloan Digital Sky Survey (SDSS). 
E, S0 and S0-a galaxies are shown respectively with
red, green and blue symbols (no sample E/S0 galaxies
have SDSS colors).
The zero-redshift ``red sequence'' region is shown in Figure 8 as a
solid line bracketed by an approximate $\pm3\sigma$ dashed lines
as found by Schawinski et al. (2009). 
Notice that S0 and S0-a galaxies appear to occupy 
approximately the same region in Figure 8 
extending from the red sequence to the blue
cloud region below, schematically shown as a shaded area
(see Baldry et al., 2004, 2006). About 15\% of early-type galaxies
in our sample are in the blue cloud (see also Schawinski et al. 2009,
Kannappan, Guie \& Baker 2009).
The lower panel of Figure 7 shows the 
corresponding infrared color-magnitude 
plot for galaxies in our sample 
and the SINGS sample ($\times$ symbols) having SDSS colors 
plotted in Figure 8. 
The red (blue) symbols in the lower panel of Figure 7 represent 
galaxies that lie above (below) the lower dashed line 
in Figure 8. 
Symbols for E galaxies in this panel are filled 
and those for S0 or S0-a are open.
What is striking about this panel is the relatively high 
specific star formation rates (larger Log$L_{24}/L_K$) for many 
galaxies that occupy the red-sequence region of Figure 8.
Conversely, several galaxies that lie below the red sequence 
region in Figure 8 are seen to have red and dead 
infrared colors in the lower panel of Figure 7, i.e. with 
Log$L_{24}/L_K$ typical of an old stellar population.

In Figure 9 we plot both samples in the  
$L_{24}/L_{70} - L_{70}/L_{160}$ color-color plot. 
Most SINGS galaxies in this plot have 
$L_{24}/L_{70} \approx -0.63 \pm 0.20$, 
resembling the $L_{24}/L_{70}$ of  
early-type galaxies having low $L_{24}/L_{70}$. 
This plot suggests that the relative lack of scatter 
along the SINGS-sample correlation  
in Figure 5 is due in part to the general similarity 
of the $L_{24}/L_{70}$ and $L_{24}/L_{160}$ colors in 
dust emission temperatures among all galaxies. 

\section{Discussion}

How closely do the galactic SFR and FIR luminosities depend on 
the mass of cold HI or H$_2$ gas? 
To answer this question we begin by plotting in Figure 10 
the observed mass of HI and H$_2$ per unit $L_{Ks}$
as a function of ${\rm Log} L_{70}/L_{Ks}$, 
duplicating (the upper panel of) Figure 2 and Figure 5 
but replacing $L_{24}$ with $M_{HI}$ and $M_{H2}$.
While correlations are clearly visible in Figure 10 -- 
approximately with 
$M_{H2}/L_K \propto L_{70}/L_K \propto M_{HI}/L_K$ -- 
the scatter with $ M_{HI}/L_K$ (Fig. 10 upper panel) 
is significantly larger than those in Figures 2 and 5.
The much lower scatter in the bottom panel 
of Figure 5 supports the widely accepted notion that 
stars form from molecular not atomic gas. 
Nevertheless, galaxies with similar $L_{24}$ can have 
masses of molecular gas that vary over an order of magnitude.
At fixed values of $L_{70}/L_K$
the vertical scatter in Figure 2 is 
slightly less than that in the lower panel of Figure 10,  
suggesting that $L_{24}/L_{Ks}$ is a better indicator 
of the specific SFR than $M_{H_2}/L_{Ks}$. 
Notice also that the SINGS galaxies occupy most of 
the same region in Figure 10 as the E and S0 galaxies in 
our sample, 
except near the upper and lower limits of the correlation.

In Figure 11 we explore this further by plotting $L_{24}$ against
the total mass of known cold gas $M_{gas} = M_{HI} + M_{H_2}$.  
It is seen that $L_{24}$ can remain unchanged when 
$M_{gas}$ varies by more than an order of magnitude 
and conversely, $L_{24}$ can vary by 10-100 among galaxies 
with the same $M_{gas}$.
The relationship between cold gas content and the SFR 
has a large scatter. 
Star-formation, as indicated by the 24$\mu$m 
emission, may be more sensitive to the local column density 
of molecular gas rather than its 
mean column density or total mass. 

In principle the mass of cold gas and 
the star formation rate SFR$(t)$ can be used 
to estimate the time required to entirely consume the cold gas.
An estimate of 
the gas consumption time is best done 
if we consider a subsample 
of S0 galaxies having similar morphologies and stellar mass.
Consider, for example, an S0 galaxy with 
$L_{Ks} \approx 3 \times 10^{10}$ $L_{Ks,\odot}$ 
and initial $L_{i,24}$ with ${\rm Log}L_{i,24}/L_{Ks} \approx 43$
near the top of Figure 1 
and suppose this galaxy
initially contains cold gas of mass $M_i = M_{gas}$.
Assume also that the galaxy forms stars at a rate SFR($L_{24}$) 
given by equation (2) based on Calzetti et al. (2007).
Setting aside the complication of mass return 
back to the interstellar medium from both young and old stars,
the time required for all the gas to be consumed into stars is
\begin{equation}
t = \int_{L_{i,24}}^{L_{f,24}} 
{\left( dM_{gas} \over d L_{24}\right)}
{d L_{24} \over {\rm SFR}(L_{24})}
~~~{\rm yr}
\end{equation}
where ${\rm Log}(L_{f,24}/L_{Ks}) \approx 41$ is the 
maximum circumstellar 
emission expected from a gas-free S0 galaxy 
with $L_{Ks} \approx 3 \times 10^{10}$ $L_{Ks,\odot}$ 
(Fig. 1).
To estimate ($dM_{gas} / d L_{24}$) 
in equation (3) we plot in Figure 12 
the observed $M_{gas}(L_{24})$ relation for galaxies having 
$10^{10} < L_{Ks} < 10^{11}$ $L_{Ks\odot}$. 
Unfortunately, it is clear from Figure 12 that 
the variation of $dM_{gas} / d L_{24}$ with $L_{24}$ 
is too uncertain to estimate the time required for the S0 
galaxy to become ``red and dead''.
Evidently either (i) uncertainties in $M_{gas}$ observations 
create too much scatter in Figure 12,
(ii) rotationally supported cold gas in S0 galaxies 
having similar $M_{gas}$ can be 
radially distributed in many different (post-merger?) ways, 
causing strong variations in local column density, the SFR  
and therefore $L_{24}$, or
(iii) star formation may occur in a very intermittent 
fashion, causing $L_{24}$ to have large variations 
for a given $M_{gas}$.
Moreover, for some well-studied S0 galaxies with low SFRs 
the rate of stellar mass loss could in principle replenish 
cold gas at approximately the same rate that it is 
consumed by star formation (TBM09).

Nevertheless, since the uncertain slopes $dM_{gas} / d L_{24}$ 
in Figures 10 and 11 appear to be 
quite large, it is possible that the 
star-formation rates in the cold gas are generally low 
and the gas-depletion times in S0 galaxies are long. 
The dusty star-forming regions in S0 galaxies 
are typically located in rotationally supported kpc-sized 
disks (e.g. 
Combes, Young \& Bureau 2007; 
Young, Bureau \& Cappellari 2008 and
Young, Bendo \& Lucero 2009).

There is general agreement that at least some cold gas disks with  
masses $\sim 10^8 - 10^9$ $M_{\odot}$ in S0 galaxies 
arise as a result of mergers, as confirmed in a few cases 
by counter-rotation (e.g. Bertola et al. 1992;
Kannappan \& Fabricant 2001;
Kannappan, Guie \& Baker 2009:
Sil'chenko, Moiseev \& Afanasiev 2009).
The statistical frequency of counter-rotating gas in 
early-type galaxies 
suggests that about half of them have acquired 
cold gas in a merger or related event 
(Kannappan \& Fabricant 2001).
To our knowledge no stars are observed to be 
counter-rotating with the cold gaseous S0 disks, 
but such observations might be possible by measuring 
the velocity of the absorption responsible for the 
H$\beta$ index.
Evidently either the mergers are very recent or 
(more likely?) the mergers are 
quite old but the star formation rates are quite low.
Although our data are sparse, 
the range of optical colors among the S0 galaxies in our sample 
is substantial, $-0.1 \lta U-V \lta 1.4$.
If blue S0s acquired their cold gas more than a few Gyrs ago,
they would be expected to exhibit a wide range of 
colors at high redshift, extending from the ``red sequence'' 
well into 
the ``blue cloud''.

\section{Conclusions}

Our main conclusions are:

\vskip.1in
\noindent(1){
S0 galaxies are sources of a broad range of mid and far infrared emission.
At the gas-free extreme, their emission at 24$\mu$m is circumstellar  
and emission at 70 and 160$\mu$m is likely to result from interstellar 
dust heated by diffuse starlight and the hot virialized interstellar 
gas (TMB07).
However, many S0 galaxies appear to contain substantial masses 
of cold gas and dust which is heated by star formation 
and radiates at all Spitzer MIPS wavelengths. 
Using estimates of the star formation rate from the 24$\mu$m luminosity 
(Calzetti et al. 2007 and Eqn. 2), 
the most infrared-luminous S0 galaxies are currently forming stars at 
1-10 $M_{\odot}$ yr$^{-1}$.
Many S0s have blue $U-V$ colors consistent with recent 
star formation. 
}

\vskip.1in
\noindent(2){
For our large sample of early-type E and S0 galaxies 
we find that 
infrared color-color diagrams 
with ${\rm Log}(L_{24}/L_{Ks})$ plotted against 
${\rm Log}(L_{70}/L_{Ks})$ or ${\rm Log}(L_{160}/L_{Ks})$
result in unusually tight correlations. 
At low values of ${\rm Log}(L_{70}/L_{Ks})$ 
or ${\rm Log}(L_{160}/L_{Ks})$ 
these plots saturate at ${\rm Log}(L_{24}/L_{Ks}) \approx 30.1$, 
corresponding to the typical SED for gas-free galaxies where 
there is no evidence for star formation and 
all the 24$\mu$m emission is circumstellar dust emission from old 
population giants.
As the mass of cold dusty gas increases in S0 galaxies 
relative to that in the stars,
${\rm Log}(L_{70}/L_{Ks})$
and ${\rm Log}(L_{160}/L_{Ks})$ increase 
and the color-color correlations are elongated in the 
direction of decreasing $L_{Ks}$.
}

\vskip.1in
\noindent(3){
Color-color plots of ${\rm Log}(L_{24}/L_{Ks})$ against
${\rm Log}(L_{70}/L_{Ks})$ or ${\rm Log}(L_{160}/L_{Ks})$ 
for E and S0 galaxies 
are coincident with the same plots for normal nearby 
(non-active) galaxies in the SINGS sample
(Kennicutt et al. 2003).
Furthermore, the SINGS galaxies are arranged along the 
infrared color-color correlations approximately following the 
Hubble morphological sequence, E $\rightarrow$ Sa $\rightarrow$
Sc $\rightarrow$ Im.
An important exception to this are the S0 galaxies which are spread
along the entire color-color correlation. 
The specific star formation rates per unit $L_{K}$ (or stellar mass) 
for some S0 galaxies is comparable to those of 
spiral or irregular galaxies that lie near them in the 
infrared color-color plot (Fig. 2).
The significant gas content in many S0 galaxies 
is likely to be relevant toward understanding the
nature and origin of these galaxies. 
}

\vskip.1in
\noindent(4){
Far-infrared luminosities correlate only
weakly with the mass of neutral and molecular hydrogen 
for SINGS galaxies and S0 galaxies from our sample.
This suggests that star formation may be intermittent
or that disks with nearly 
equal masses of cold gas are distributed differently with 
galactic radius
in similar S0 galaxies so that the variation of local cold gas (volume
and column) densities alters the local and global star formation
rates.
}

\vskip.1in
\noindent(5){
Early type galaxies do not have infrared properties that 
strictly adhere to the red sequence and blue cloud 
regions of the SDSS color magnitude plot.
For example, in an IR color-magnitude plot 
quite a few galaxies in the SDSS red sequence 
occupy an infrared blue cloud region 
with larger Log$L_{24}/L_K$ 
than passively evolving ellipticals, 
indicating relatively high specific star formation rates. 
Conversely, a few galaxies 
occupying the green valley or blue cloud region of 
the SDSS color magnitude plot have infrared star formation rates 
that are indistinguishable from pure gas-free, passively-evolving old 
stellar populations that define the infrared red sequence.
}

\vskip.1in
\acknowledgements
This work is based on observations made with the Spitzer
Space Telescope, which is operated by the Jet Propulsion
Laboratory, California Institute of Technology, under NASA
contract 1407. This publication makes use of data products from the Two
Micron All Sky Survey - which is a joint project of the University of
Massachusetts and the Infrared Processing and Analysis Center/California
Institute of Technology - and  data from the Sloan Digital Sky Survey
(SDSS).
We acknowledge the usage of the HyperLeda database
(http://leda.univ-lyon1.fr) and
the NASA/IPAC Extragalactic Database (NED).
Support for this work was provided by NASA
through Spitzer Guest Observer grant RSA 1276023.
Studies of the evolution of hot gas in elliptical galaxies
at UC Santa Cruz are supported by NSF and NASA grants
for which we are very grateful.


\clearpage

                                                                                   
\clearpage

\begin{figure}
\centering
\includegraphics[bb=200 100 422 469,scale=0.8,angle=0]{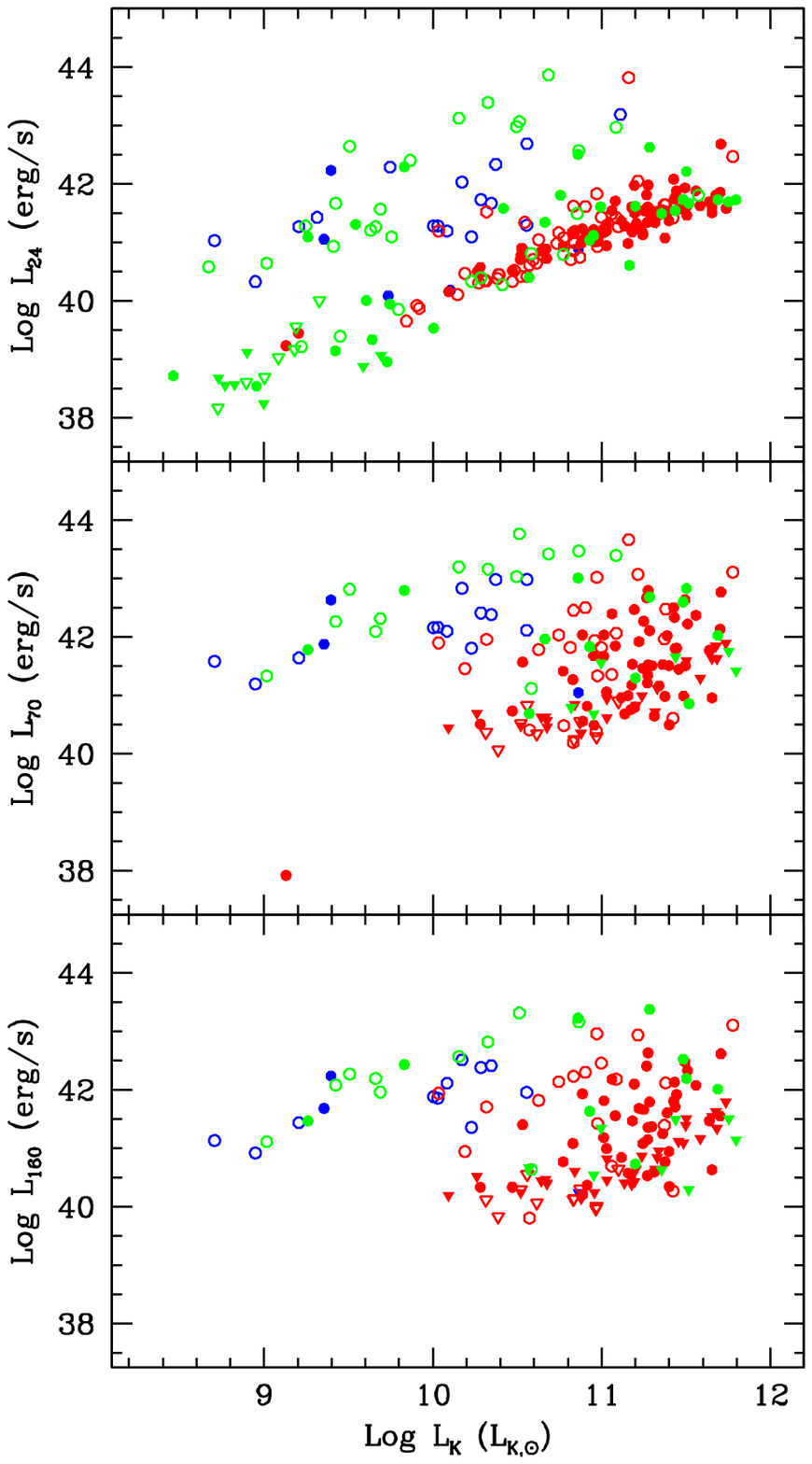}
\vskip.7in
\caption{
Spitzer data for elliptical and S0 galaxies. 
Circles represent observations, triangles upper limits.
Filled symbols refer to elliptical 
galaxies with de Vaucouleurs classification 
parameter $T < -3$; open symbols refer to S0 galaxies with $T > -3$.
Red (blue) symbols have optical colors $U-V < 1.1$ ($U-V > 1.1$);
galaxies with unknown colors are plotted with green symbols. 
}
\label{fig1}
\end{figure}

\clearpage

\begin{figure}
\centering
\includegraphics[bb=180 100 422 469,scale=0.8,angle=0]{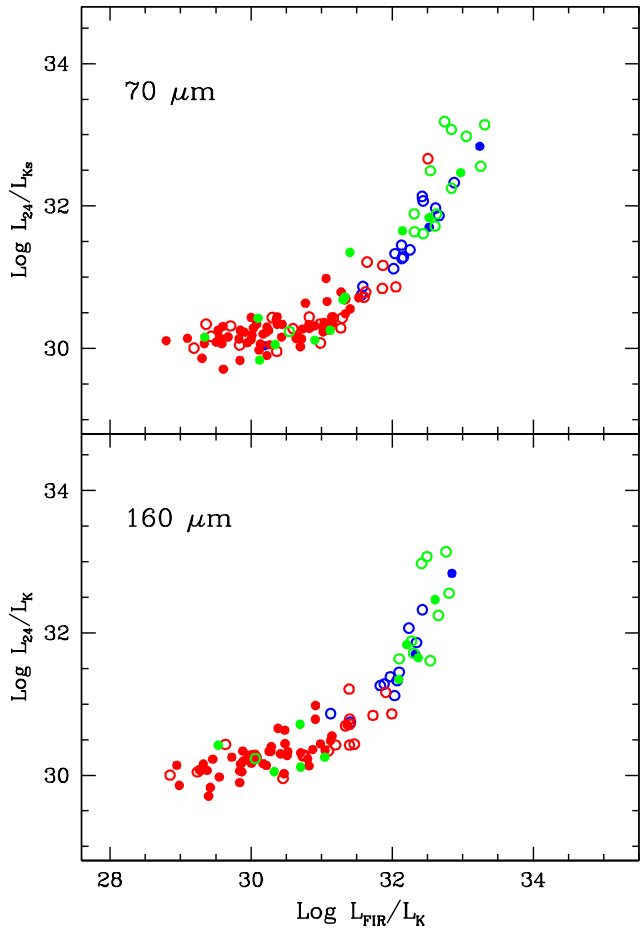}
\vskip.7in
\caption{
Infrared color-color plots of $L_{24}/L_{Ks}$ against
$L_{70}/L_{Ks}$ (upper panel) and $L_{160}/L_{Ks}$ (lower panel) 
for E (filled circles) and S0 (open circles) galaxies. 
Galaxies with red symbols have colors $U-V > 1.1$ 
indicating less star formation than those with blue 
symbols ($U-V < 1.1$). 
Green symbols identify galaxies 
for which optical colors are unavailable.
}
\label{fig2}
\end{figure}

\clearpage

\begin{figure}
\centering
\includegraphics[bb=200 116 422 469,scale=0.80,angle=0]{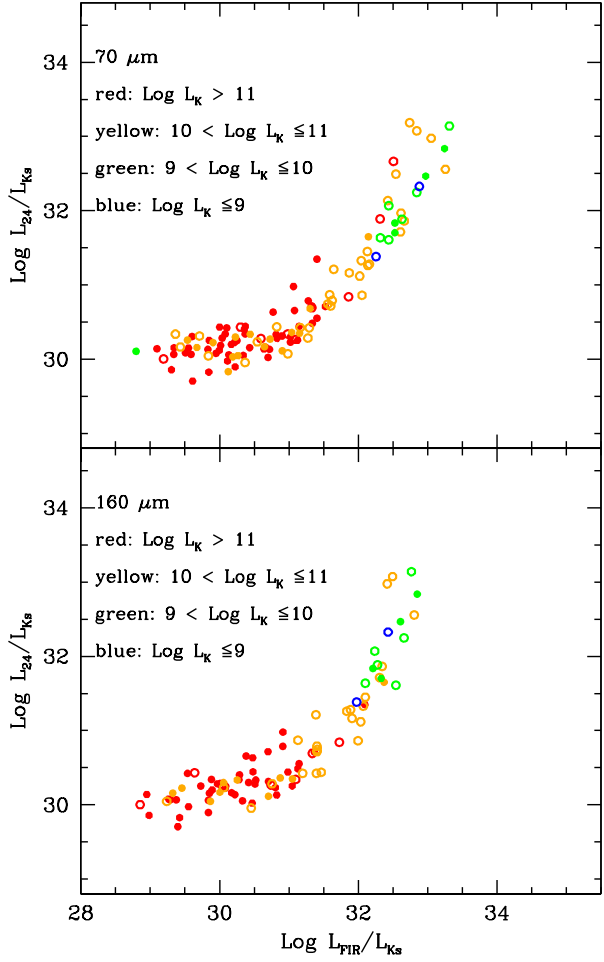}
\vskip.7in
\caption{
Infrared color-color plots of $L_{24}/L_{Ks}$ against
$L_{70}/L_{Ks}$ and $L_{160}/L_{Ks}$
for E (filled circles) and S0 (open circles) galaxies. 
Galaxies are color-binned with increasing $L_{Ks}$.
The green galaxy to the far left in the upper panel 
is NGC 221 (M32).
}
\label{fig3}
\end{figure}

\clearpage

\begin{figure}
\figurenum{4a}
\centering
\includegraphics[bb=180 116 422 469,scale=1.01,angle=0]{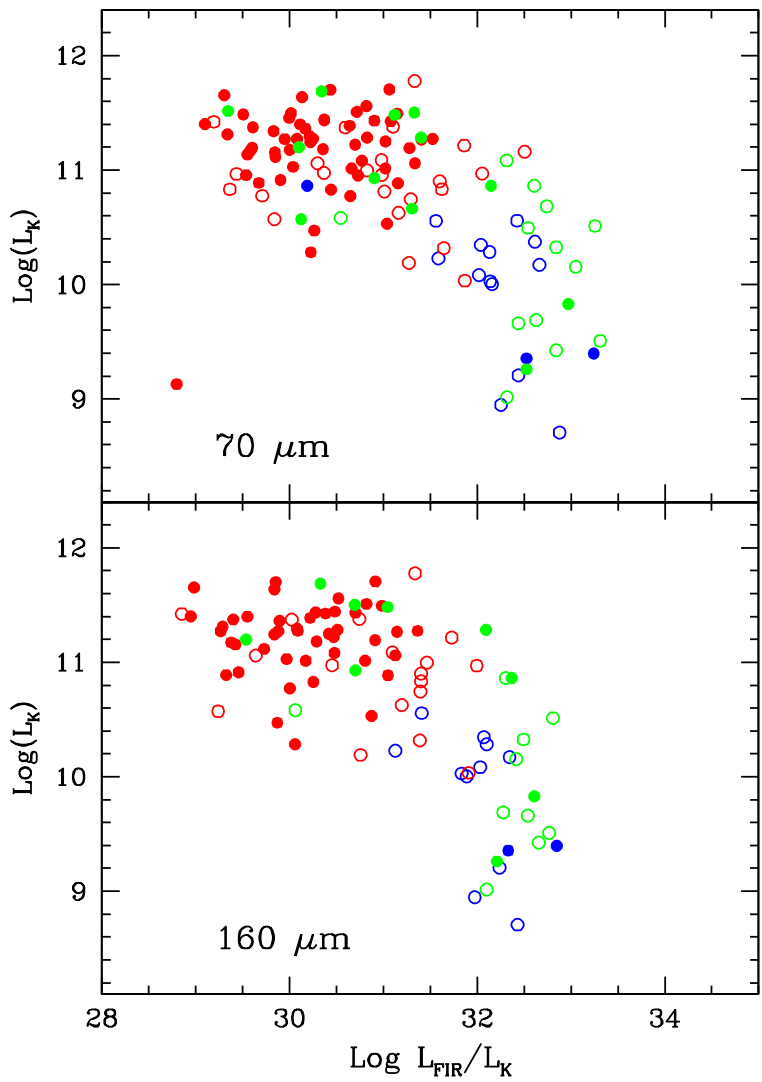}
\vskip.7in
\caption{
Variation of $L_{Ks}$ with 
${\rm Log} L_{FIR}/L_{Ks}$, the horizontal axis in Figure 2,
where FIR represents 70 (upper panel) or 160$\mu$m (lower panel).
Colors and symbols are identical to those in Figure 2.
The red galaxy at the lower left in the upper panel
is NGC 221 (M32).
}
\label{fig4a}
\end{figure}

\clearpage

\begin{figure}
\figurenum{4b}
\centering
\includegraphics[bb=180 116 422 469,scale=1.01,angle=0]{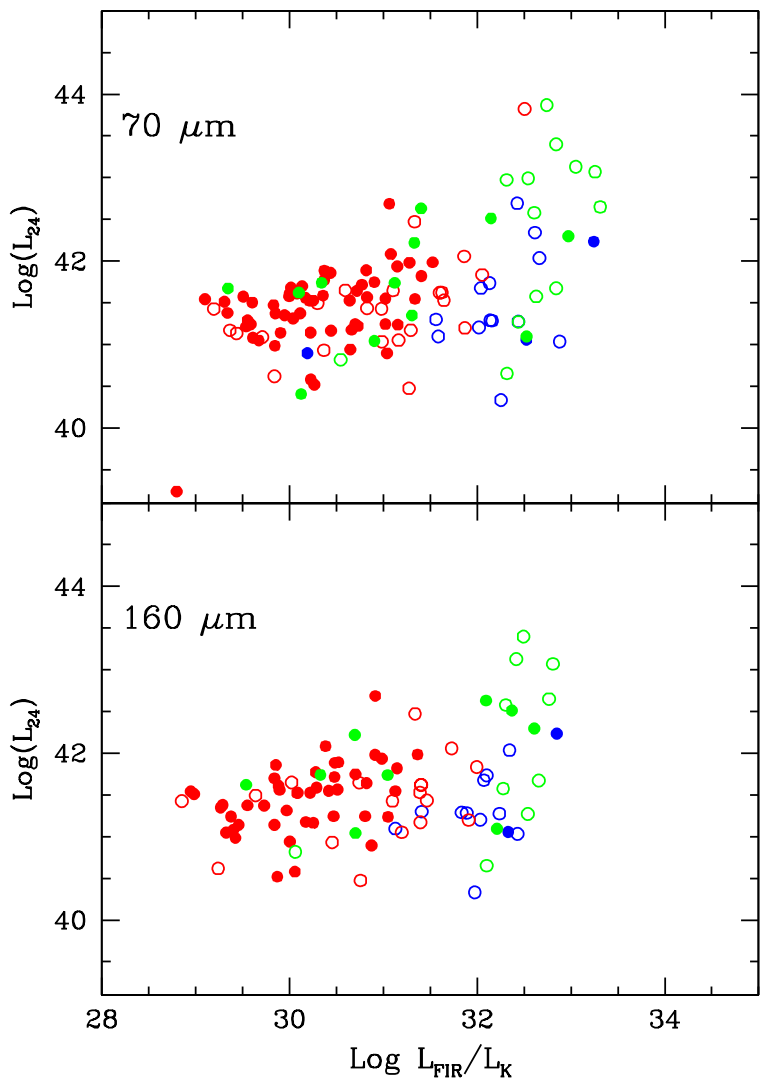}
\vskip.7in
\caption{
Variation of $L_{24}$ with
${\rm Log} L_{FIR}/L_{Ks}$, the horizontal axis in Figure 2.
Colors and symbols are identical to those in Figure 2.
The red galaxy at the lower left in the upper panel
is NGC 221 (M32).
}
\label{fig4b}
\end{figure}

\clearpage

\begin{figure}
\figurenum{5}
\centering
\includegraphics[bb=170 116 422 469,scale=1.01,angle=0]{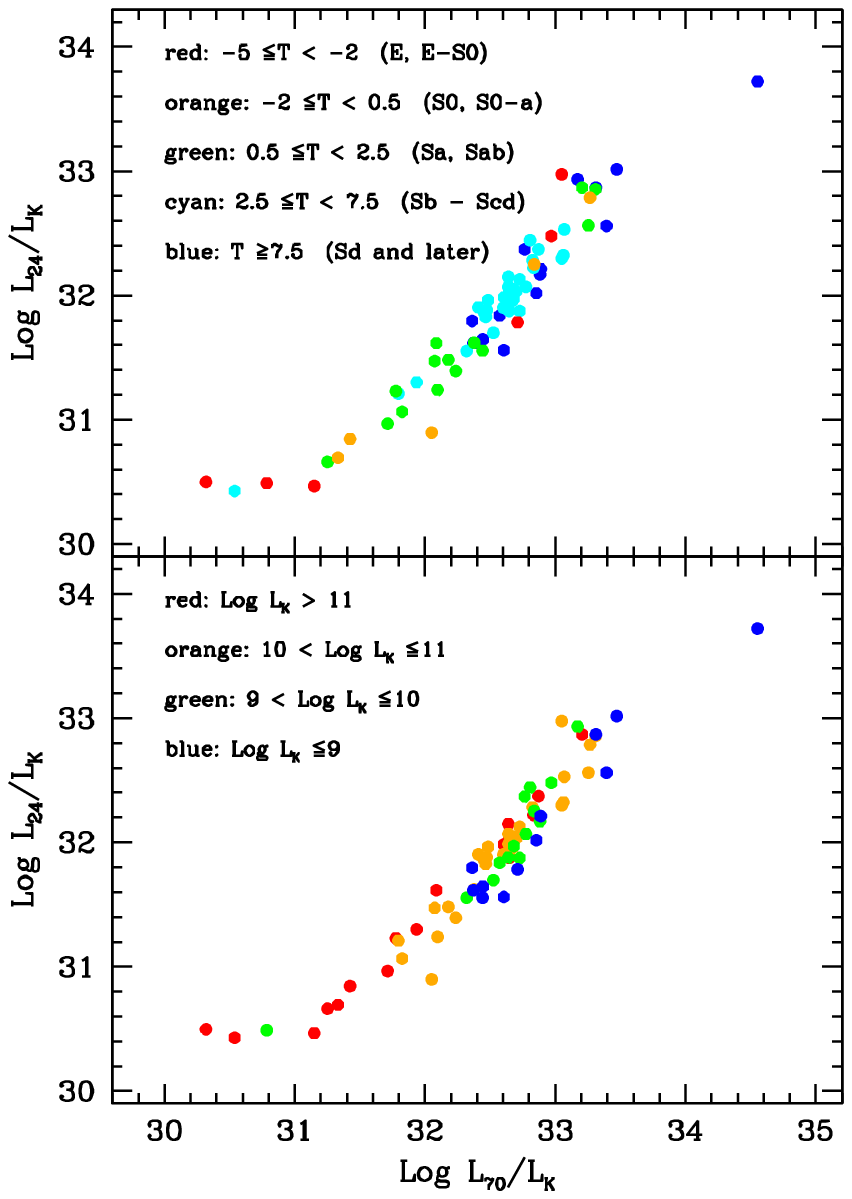}
\vskip.7in
\caption{
Spitzer data for SINGS galaxies. 
{\it Upper panel:} Color-binned by morphological T 
parameter.
{\it Lower panel:} Color-inned in $L_{Ks}$. 
}
\label{fig5}
\end{figure}

\clearpage

\begin{figure}
\figurenum{6}
\centering
\includegraphics[bb=170 216 422 469,scale=1.01,angle=0]{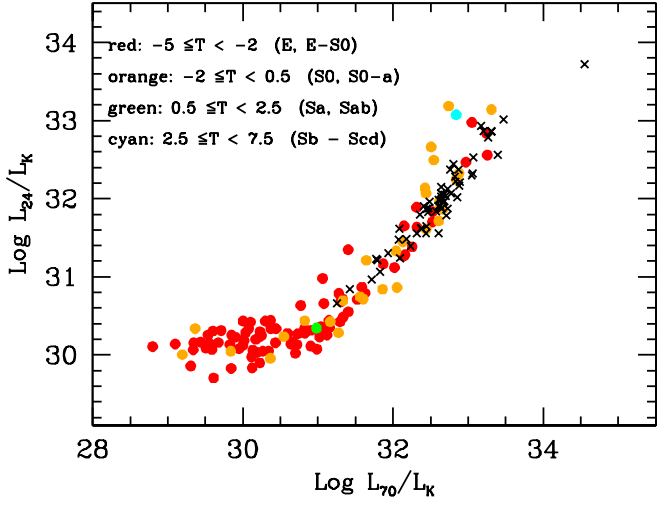}
\vskip.7in
\caption{
Infrared color-color plot for our sample of 
early-type galaxies combined with data 
for normal SINGS galaxies (not already in our sample) 
represented with black 
$\times$ symbols. 
}
\label{fig6}
\end{figure}

\clearpage

\begin{figure}
\figurenum{7}
\centering
\includegraphics[bb=200 156 422 469,scale=0.80,angle=0]{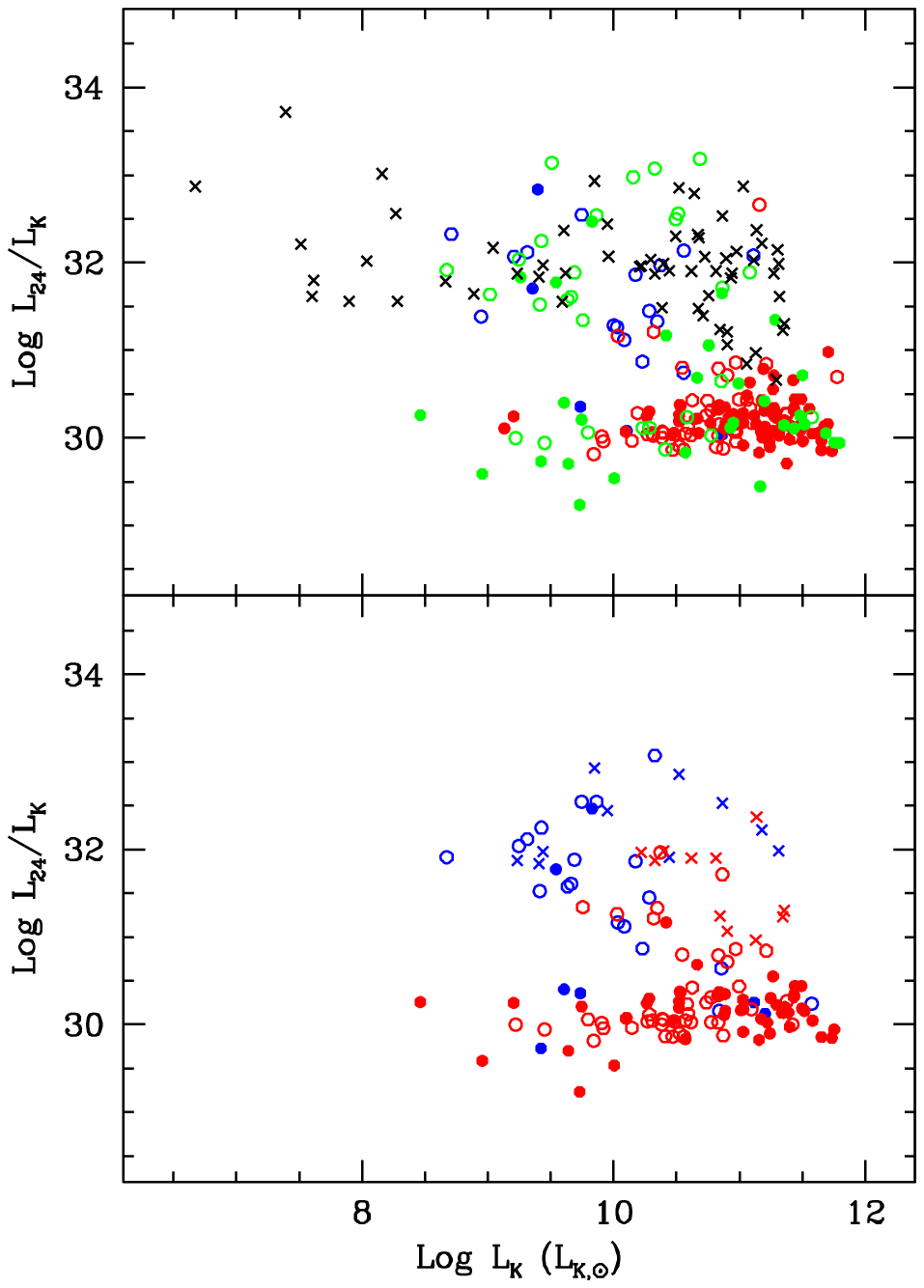}
\vskip.7in
\caption{
{\it Upper panel:} Combined plot of Spitzer data of 
early type galaxies from our sample and SINGS galaxies 
of all morphologies.
Filled circles refer to elliptical
galaxies with de Vaucouleurs classification
parameter $T < -3$; open circless refer to S0 galaxies with $T > -3$.
Red (blue) symbols have optical colors $U-V < 1.1$ ($U-V > 1.1$), and 
galaxies with unknown colors are not plotted. 
SINGS galaxies are represented with 
black $\times$ symbols regardless of morphology. 
{\it lower panel:} A similar plot showing early-type 
galaxies from our sample together with the SINGS sample.  
Symbols are as in the upper panel except now 
red symbols indicate that the galaxies in our sample 
have $u-r$ colors that 
lie within 3$\sigma$ of the zero-redshift SDSS red sequence 
while blue symbols correspond to galaxies with bluer 
$u-r$ colors.
}
\label{fig7}
\end{figure}

\clearpage

\begin{figure}
\figurenum{8}
\centering
\includegraphics[bb=160 200 442 469,scale=1.01,angle=0]{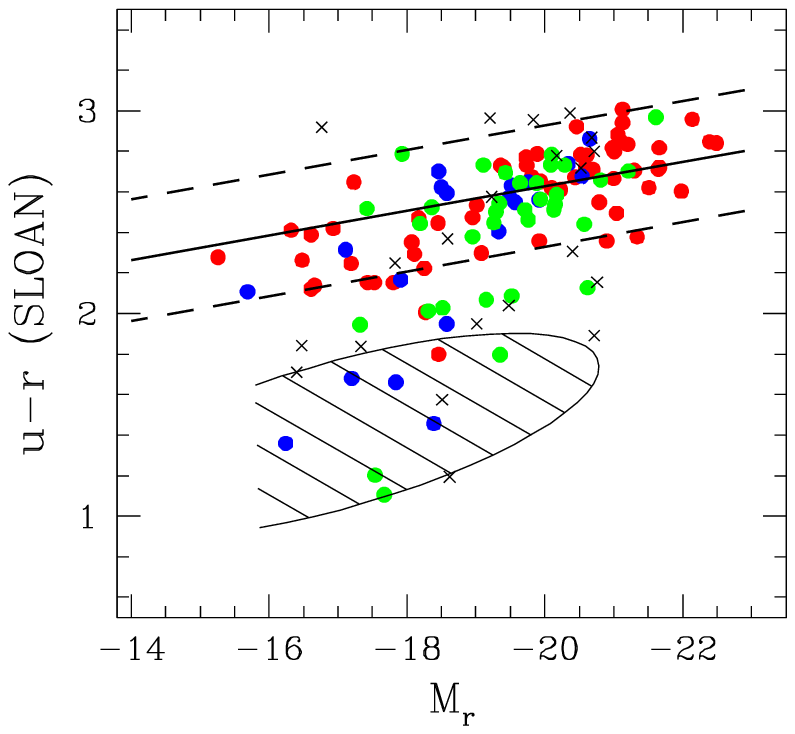}
\vskip.7in
\caption{
SDSS color-magnitude plot showing galaxies from our 
early-type sample and the SINGS sample.
Only galaxies with SDSS colors are plotted.
Red symbols are E galaxies (no sample E/S0 galaxies have SDSS 
colors), green are S0 galaxies and blue are S0-a galaxies. 
SINGS galaxies are represented with
black $\times$ symbols regardless of morphology.
The solid line shows the zero-redshift red sequence defined 
as an approximate $\pm3\sigma$ region bracketed with dashed lines.
An approximate zero-redshift blue cloud region is shown below 
(shaded region).
}
\label{fig8}
\end{figure}

\clearpage

\clearpage

\begin{figure}
\figurenum{9}
\centering
\includegraphics[bb=160 180 442 469,scale=1.01,angle=0]{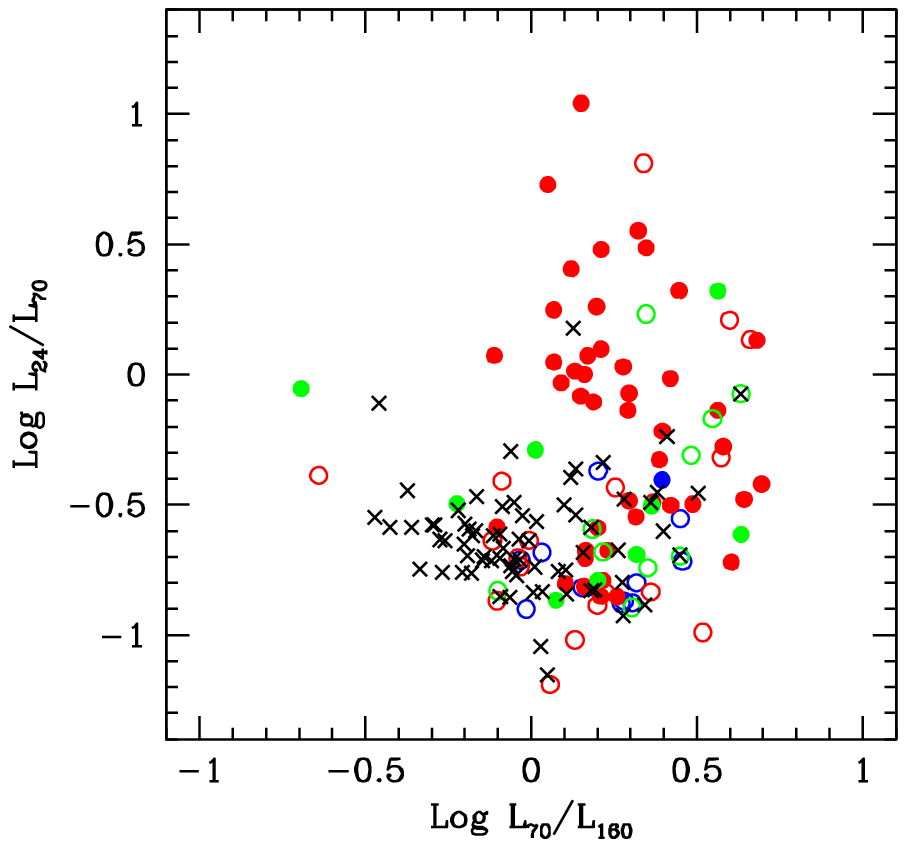}
\vskip.7in
\caption{
Combined color-color plot of
early-type and SINGS galaxies (symbols as in Figure 7).
}
\label{fig9}
\end{figure}

\clearpage

\begin{figure}
\figurenum{10}
\centering
\includegraphics[bb=170 150 422 469,scale=1.01,angle=0]{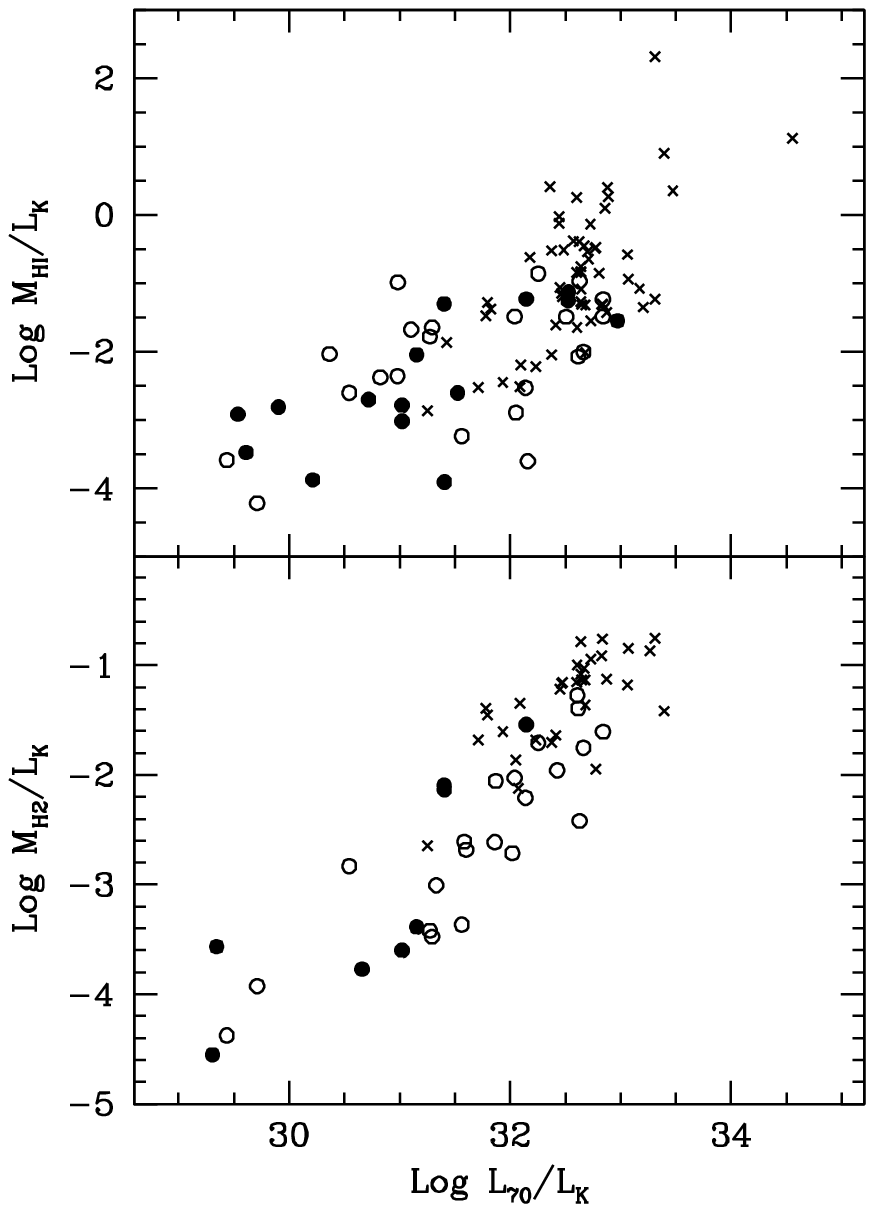}
\vskip.7in
\caption{
Plots of the mass of neutral and molecular cold gas per unit 
$L_{Ks}$ against 
${\rm Log} L_{70}/L_{Ks}$, the horizontal axis in Figure 2 
and Figure 5.
Galaxies from our early type sample are designated with 
filled (E) or open (S0) circles; galaxies from the SINGS sample 
are shown with $\times$ symbols.
}
\label{fig10}
\end{figure}

\clearpage

\begin{figure}
\figurenum{11}
\centering
\includegraphics[bb=150 290 422 569,scale=.8,angle=0]{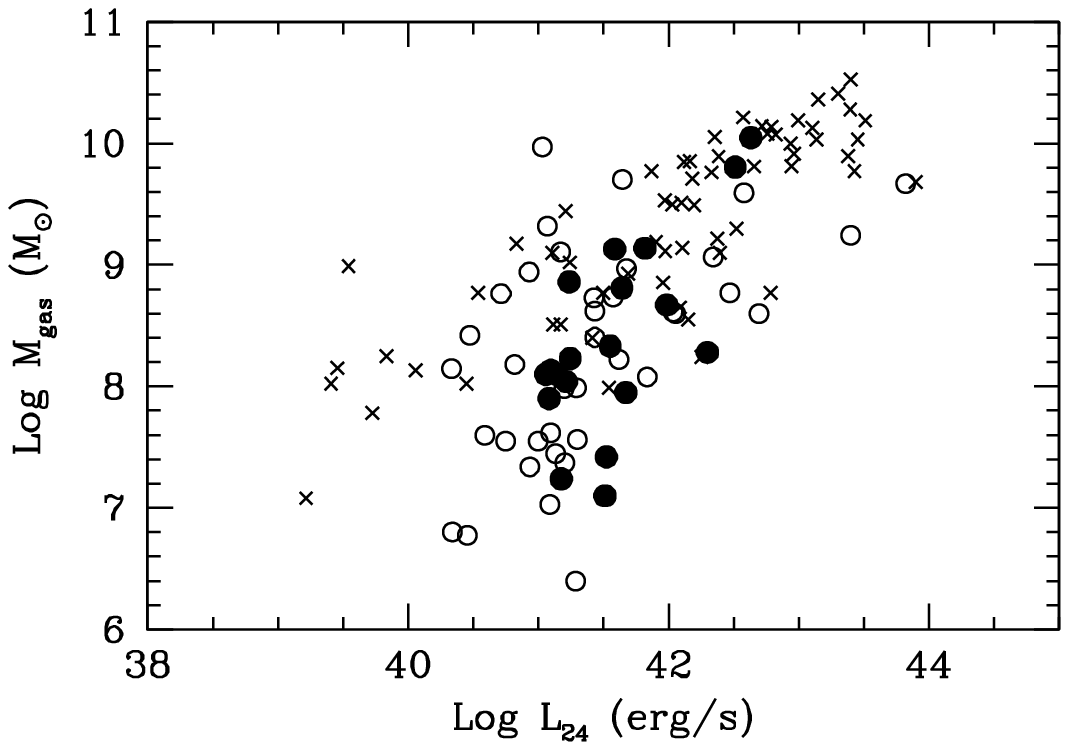}
\vskip.7in
\caption{
Plot of the total observed mass of cold gas 
$M_{gas} = M_{HI} + M_{H_2}$ against the 24$\mu$m luminosity.
Galaxies from our early type sample are designated with
filled (E) or open (S0) circles; galaxies from the SINGS sample
are shown with $\times$ symbols.
}
\label{fig11}
\end{figure}

\clearpage

\begin{figure}
\figurenum{12}
\centering
\includegraphics[bb=150 290 422 569,scale=.8,angle=0]{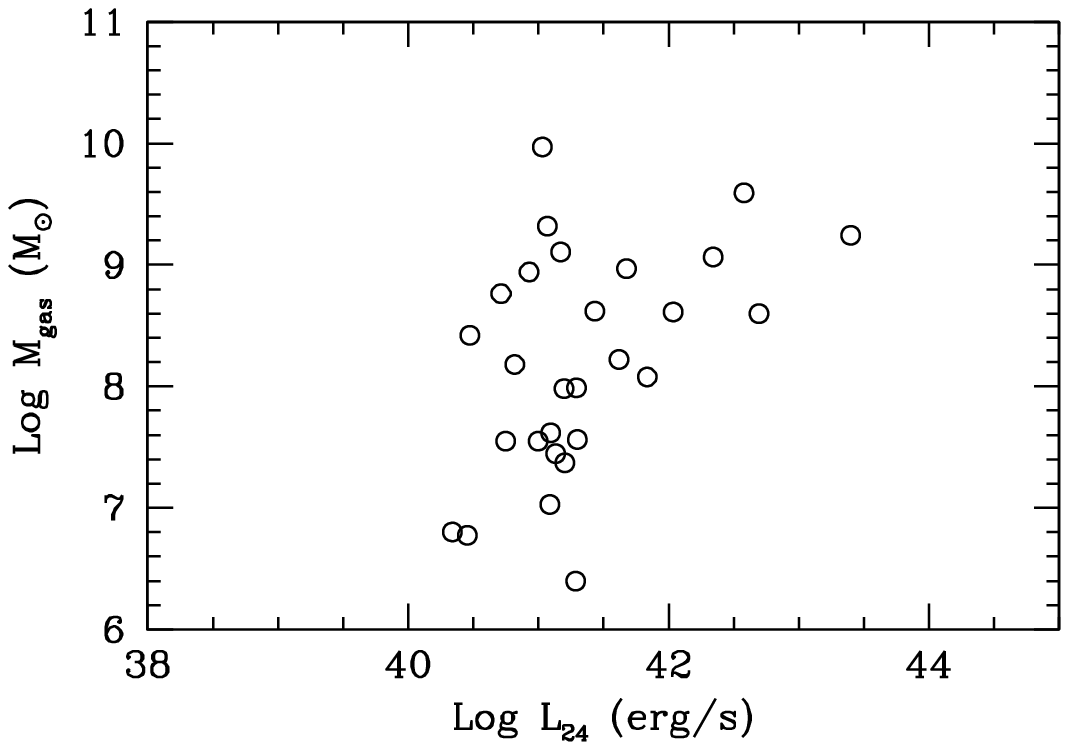}
\vskip.7in
\caption{
Similar to Figure 11, but restricted to S0 galaxies having
$10 < {\rm Log}L_{Ks} < 11$.
}
\label{fig12}
\end{figure}


\clearpage

\begin{deluxetable}{lllrrrrrrrrrrrr}
\tabletypesize{\tiny}
\rotate
\tablecaption{Early Type Galaxies from the Spitzer Archives}
\tablecaption{}
\tablewidth{0pt}
\tablehead{
\colhead{Name} & \colhead{PI/ID\#} & \colhead{Type} & \colhead{T} & \colhead{D} &
\colhead{Log $L_{Ks}$} & \colhead{U-V } & \colhead{Log $M_{HI}$} & \colhead{Log $M_{H_2}$} & \colhead{$F_{24\mu m}$}& \colhead{$F_{70\mu m}$}&
\colhead{$F_{160\mu m}$} & \colhead{Log$L_{24\mu m}$} &\colhead{Log$L_{70\mu m}$} &\colhead{Log$L_{160\mu m}$} \\
\colhead{} & & \colhead{} & \colhead{} &
\colhead{(Mpc)} &\colhead{($L_{Ks,\odot}$)}&&\colhead{($M_\odot$)} &\colhead{($M_\odot$)} &\colhead{(mJy)} &\colhead{(mJy)}&\colhead{(mJy)}&
\colhead{$(erg \ s^{-1}$)} & \colhead{$(erg \ s^{-1}$)} &\colhead{$(erg \ s^{-1}$)} \\
\colhead{(1)} & \colhead{(2)} & \colhead{(3)} & \colhead{(4)} & \colhead{(5)} & \colhead{(6)} & \colhead{(7)} &
\colhead{(8)} & \colhead{(9)} & \colhead{(10)} & \colhead{(11)} & \colhead{(12)} & \colhead{(13)} & \colhead{(14)} & \colhead{(15)} \\
}
\startdata
     
    N0221 & Fazio/69 & E & -4.7 &   0.81 & 9.13 & 1.140& \nodata& \nodata& 175$\pm$16&25$\pm$18&\nodata &39.24&37.93&\nodata \\
    N0315 & Fazio/69 & E & -4.0 &  58.88 & 13.44 & 1.487& \nodata &\nodata& 94$\pm$6&331$\pm$17&536$\pm$48 &42.69&42.77&42.62 \\
    N0404 & Fazio/69 & E-S0 & -2.8 &   3.27 &   8.95 & 1.095& 8.09 & 7.24 &135$\pm$8&2919$\pm$202&3499$\pm$366 &40.33&41.20&40.92 \\
    N0410 & Fazio/69 & E & -4.3 &  56.75 & 11.50 & 1.459 &\nodata&\nodata& 6.1$\pm$0.8&$\leq24$&$\leq45$ &41.47&$\leq41.60$&$\leq41.51$ \\
    N0474 & Zezas/20140 & S0 & -2.0 &  33.46 & 10.98 & 1.157 &8.94 & 7.63&5.1$\pm$2.2&38.2$\pm$7.5&107$\pm$22 &40.93&41.34&41.43 \\ 
    N0507 & Temi/20171  & E-S0 & -3.3 &  67.19 & 11.68 & 1.364 &\nodata &\nodata& \nodata&$\leq18.8$&$\leq22.1$ &\nodata&$\leq41.64$&$\leq41.35$ \\
    N0516 & Kannappan/30406 & S0-a & -1.5 &  35.01 &  10.22 & \nodata & \nodata  &\nodata& 1.2$\pm$0.3&\nodata&\nodata &40.34&\nodata&\nodata \\
    N0526 & Werner/86 & S0 & -2.0 &  78.70 & 3.14 & \nodata & \nodata  &\nodata& 307$\pm$28&294$\pm$41&\nodata &43.45&42.97&\nodata \\
    N0533 & Temi/20171 & E & -4.8 &  68.23 & 11.64 & 1.495 & \nodata   &\nodata& 7.2$\pm$3.1&24.8$\pm$4.9&28.7$\pm$5.6 &41.70&41.77&41.48 \\
    N0584 & Kennicutt/159 & E & -4.6 &  23.76 & 11.18 & 1.358 & \nodata   &\nodata& 48.5$\pm$7.6&52.3$\pm$6.9&$\leq18.9$ &41.61&41.18&$\leq40.38$ \\
    N0596 & Surace/3403 & E & -4.3 &  23.87 & 10.91 & 1.239 & 8.1  &\nodata& 16.2$\pm$4.6&22.5$\pm$5.1&18.4$\pm$7.3 &41.14&40.82&40.37 \\
    N0636 & Surace/3403 & E & -4.8 &  22.28 & 10.67 & 1.372 & \nodata & 7.15  &10.5$\pm$3.6&$\leq17.2$&$\leq26.9$ &40.89&$\leq40.64$&$\leq40.48$ \\
    N0720 & Temi/20171 & E & -4.8   & 22.29 & 11.14 & 1.410 & \nodata  & 7.46 &26.1$\pm$5.1&19.2$\pm$4.9&$\leq23.1$ &41.29&40.69&$\leq40.41$ \\
    N0777 & Fazio/69 &E & -4.8 &   55.21 & 11.49 & 1.524 & \nodata &\nodata& 8.2$\pm$2.7&6.3$\pm$2.8&$\leq18.4$ &41.57&40.99&$\leq41.10$ \\
    N0807 & Young/20780 &E & -4.8  &  69.18   & 11.28 & \nodata & 9.98&9.19 &59.9$\pm$3.2&198$\pm$9&2233$\pm$235 &42.63&42.69&43.38 \\
    N0814 & Roussel/20577 & S0&  -1.8 &   20.70  & 9.51 & \nodata & \nodata &\nodata& 697$\pm$42&3007$\pm$265&1952$\pm$182 &42.65&42.82&42.27 \\
    N0821 & Fabbiano/20371 & E & -4.8 &   24.09  & 10.95 & \nodata & \nodata &\nodata& 15.4$\pm$4.8&$\leq16.3$&$\leq27.3$ &41.13&$\leq40.69$&$\leq40.55$ \\
    N0855 & Kennicutt/159 & E & -4.6 &    9.33 & 9.35 & 0.613 & 8.10   &\nodata& 87.8$\pm$8.5&1700$\pm$136&2486$\pm$246 &41.06&41.88&41.69 \\
    N1016 & Fazio/30318 &E&  -4.8 &   73.79 & 11.65 & 1.479 & \nodata   &\nodata& 5.1$\pm$2.9&$\leq15.4$&$\leq28.5$ &41.62&$\leq41.63$&$\leq41.54$ \\
    N1023 & Fazio/69 & E-S0 & -2.7  &  11.42  & 10.97 & 1.387 & 9.32 & 6.69 &60.1$\pm$5.6&$\leq29.6$&$\leq34.8$ &41.07&$\leq40.30$&$\leq40.01$ \\
    N1199 & Johnson/3596 & E&  -4.6 &   32.6  &  10.95 & 1.346 & \nodata  &\nodata& 10.5$\pm$1.2&88$\pm$13&\nodata &41.22&41.68&\nodata \\
    N1266 & Kennicutt/40204 & S0 & -2.1 &   30.06    & 10.51 & \nodata & \nodata &\nodata& 872$\pm$32&12690$\pm$535&10300$\pm$843 &43.07&43.77&43.32 \\
    N1316 & Kennicutt/159 &S0&  -1.8  &  21.48 & 11.78 & 1.183 & \nodata & 8.77 &430$\pm$21 &5440$\pm$312&12610$\pm$886 &42.47&43.11&43.11 \\
    N1374 & Fazio/30318 &E&  -4.3 &   19.77&  10.68 & 1.347 & \nodata   &\nodata& 9.2$\pm$1.4&$\leq18.5$&$\leq28$ &40.73&$\leq40.57$&$\leq40.39$ \\
    N1377 & Kennicutt/159&  S0 & -2.1 &    22.20&  10.15 & \nodata & \nodata    &\nodata& 1835$\pm$84&6350$\pm$371&3380$\pm$248 &43.13&43.20&42.57 \\
    N1386 & Werner/86&  S0-a & -0.8&    16.52  & 10.56 & 1.038 & \nodata& 8.60 &1211$\pm$66&6900$\pm$412&\nodata &42.69&42.98&\nodata \\
    N1395 & Kaneda/3619 & E & -4.8  &  21.98  &  11.27 & 1.486 & \nodata& \nodata& 46.4$\pm$6.5&135$\pm$11&213$\pm$14 &41.52&41.52&41.36 \\
    N1399 & Temi/20171&  E & -4.5 &   19.40   &  11.40 & 1.429 & \nodata  &\nodata& 61.9$\pm$7.3&16.4$\pm$7.6&26.5$\pm$8.6 &41.54&40.50&40.35 \\
    N1404 & Kennicutt/159&  E&  -4.7 &   19.40 &  11.20 & 1.499 &\nodata  &\nodata& 56.6$\pm$6.3&32.6$\pm$6.3&$\leq32.2$ &41.50&40.80&$\leq40.43$ \\
    N1407 & Kaneda/3619  &E&  -4.5 &   22.08  & 11.36 & \nodata & \nodata& 7.55 &43.4$\pm$6.4&$\leq21.1 $&$\leq38.9$ &41.50&$\leq40.74$&$\leq40.63$ \\
    N1426 & Fazio/30318 &E&  -4.6 &  24.10  & 10.64 & 1.292 & \nodata  &\nodata& 8.3$\pm$2.1&$\leq14.4$&$\leq21.4$ &40.86&$\leq40.63$&$\leq40.44$ \\
    N1427 & Fazio/30318 &E & -4.0&   23.55  & 2.09 & 1.300 & \nodata &\nodata& 9.5$\pm$1.4&27.3$\pm$1.1&$\leq13.3$ &40.90&40.89&$\leq40.22$ \\
    N1439 & Fazio/30318 &E & -4.7 &  26.67 & 10.77 & 1.207 & \nodata &\nodata& 8.2$\pm$0.8&72$\pm$6.2&37.5$\pm$8.0 &40.94&41.42&40.78 \\
    N1510 & Kennicutt/40204&  S0-a & -2.0  & 10.01   & 9.21 & 0.223 & \nodata  &\nodata& 126$\pm$13&862$\pm$35&1237$\pm$103 &41.28&41.65&41.44 \\
    N1522 & Kennicutt/40204 & S0  &-1.8&   9.04 &  8.71 & -0.024 & \nodata &\nodata& 88$\pm$4&924$\pm$63&749$\pm$66 &41.03&41.59&41.14 \\
    N1533 & Putman/20695 & S0 & -2.4 & 21.38  & 10.96 & 1.403 & 9.97 &\nodata& 15.8$\pm$2.3&374$\pm$18&\nodata &41.03&41.94&\nodata \\
    N1543 & Fisher/30496 & S0 & -1.9 &20.04  &  10.97 & 1.320 & \nodata &\nodata& 18.2$\pm$2.9&\nodata&\nodata &41.04&\nodata&\nodata \\
    N1553 & Fabbiano/20371&  S0 & -2.3 & 18.54  &  11.37 & 1.301 & \nodata &\nodata& 86.7$\pm$8.9&527$\pm$24&322$\pm$21 &41.65&41.97&41.39 \\
    N1700 & Surace/3403&  E&  -4.7 &  38.04  &11.27 & 1.368 &\nodata &\nodata& 19.0$\pm$4.9&30.4$\pm$5.3&44.1$\pm$7.2 &41.61&41.35&41.16 \\
    N2110 & Werner/86&  E-S0 & -3.0&  31.33   & 11.08 & \nodata & \nodata &\nodata& 641$\pm$32&4966$\pm$422&\nodata&42.97&43.40&\nodata \\
    N2300 & Zezas/20140&  E-S0 & -3.4&  27.67  &11.18 & 1.571 & \nodata &\nodata& 13.3$\pm$2.1&\nodata&\nodata &41.18&\nodata&\nodata \\
    N2325 & Temi/20171 & E & -4.7&   31.92   &  11.20 & \nodata & \nodata &\nodata& 27.5$\pm$4.4&38.3$\pm$6.8&23.9$\pm$6.4 &41.62&41.30&40.74 \\
    N2434 & Fazio/30318 &E&  -4.8&  21.58   & 10.86 & 1.090 & \nodata  &\nodata& 11.3$\pm$0.4&47.3$\pm$6.9&$\leq16.8$ &40.90&41.05&$\leq40.24$ \\
    N2685 & Rieke/40936 & S0-a & -1.1 & 15.63  &   10.39 & 1.029 & 9.26 & 7.50 &\nodata&\nodata&\nodata &\nodata&\nodata&\nodata \\
    N2768 & Temi/20171 & E&  -4.3&   22.38   & 11.25 & 1.333 &  8.23  & 7.65 &47.5$\pm$5.2&728$\pm$26&414$\pm$29 &41.55&42.27&41.67 \\
    N2778 & Fazio/30318 &E & -4.7  & 22.91 &  10.26 & 1.378 & \nodata &\nodata& 4.1$\pm$1.3&$\leq18.4$&$\leq28.2$ &40.51&$\leq40.69$&$\leq40.52$ \\
    N2787 & Fisher/30496 & S0-a & -1.1  & 7.48   &  10.19 & 1.396 & 8.41 & 6.77 &36$\pm$5&1017$\pm$46&705$\pm$48 &40.47&41.46&40.95 \\
    N2832 & Fazio/69&  E&  -4.3 &  85.90  &  11.74 & 1.447 & \nodata &\nodata& 3.5$\pm$0.5    &$\leq21$&$\leq38$ &41.59&$\leq41.90$&$\leq41.80$ \\
    N2970 & Zezas/20140 & E&  -4.6&   25.23    & 9.74 & 0.868 & \nodata &\nodata& 1.3$\pm$0.2&\nodata&\nodata &40.09&\nodata&\nodata \\
    N2974 & Kaneda/3619&  E&  -4.7&   21.48 &  11.51 & 1.453 &8.81 & 7.59 &63.7$\pm$4.4&716$\pm$48&2076$\pm$154 &41.64&42.23&42.33 \\
    N2986 & Temi/20171 & E&  -4.7 &   32.51  & 11.32 & 1.415 & \nodata &\nodata& 13.9$\pm$3.8&$\leq9.8$&$\leq18.5$ &41.34&$\leq40.72$&$\leq40.64$ \\
    N3011 & Kannappan/30406 & S0 & -1.6 &  23.77 &  9.63 & \nodata & \nodata     &\nodata& 19.2$\pm$3.2&\nodata&\nodata &41.21&\nodata&\nodata \\
    N3032 & Young/20780 & S0 & -1.8  & 21.98   &   10.17 & 0.728 &8.17  & 8.42 &151$\pm$8&2772$\pm$184&3044$\pm$273 &42.04&42.84&42.52 \\
    N3073 & Fazio/69&  E-S0 & -2.8 &  33.73  &  10.08 & 0.691 & \nodata & 7.37 &9.4$\pm$1.5&218$\pm$8&515$\pm$42 &41.20&42.10&42.12 \\
    N3115 & Fazio/69 & E-S0 & -2.9  &  9.68  &   10.97 & 1.375 & 7.38 & 6.48 &97$\pm$4&52$\pm$4&$\leq45$ &41.13&40.40&$\leq39.98$ \\
    N3125 & Rieke/40936 & E&  -4.8&  13.43 &   9.40 & -0.118 & \nodata &\nodata& 636$\pm$32&4714$\pm$273&4337$\pm$327 &42.23&42.64&42.24 \\
    N3156 & Surace/3403 & S0&  -2.4&   22.39  &  10.23 & 0.924 & \nodata & 7.62 &16.7$\pm$4.2&254$\pm$18&203$\pm$21 &41.10&41.81&41.36 \\
    N3226 & Appleton/1054 & E & -4.8 & 23.55&   10.66 & \nodata & \nodata &\nodata& 26.9$\pm$3.2&327$\pm$14&\nodata &41.35&41.97&\nodata \\
    N3265 & Kennicutt/159 & E & -4.8 &  21.33 &   9.83 & \nodata &8.28    &\nodata& 292$\pm$17&2719$\pm$144&2692$\pm$159 &42.30&42.80&42.44 \\
    N3377 & Fazio/69&  E & -4.8 & 11.22  &   10.47 & 1.102 & \nodata & 6.88 &17.6$\pm$5.4&84.6$\pm$5.7&77.7$\pm$9.2 &40.52&40.74&40.34 \\
    N3379 & Fazio/69 & E & -4.8 & 10.57 & 10.89 & 1.436 & \nodata & 6.69 &66.8$\pm$8.8&63.5$\pm$8.0&65.2$\pm$8.2 &41.05&40.56&40.21 \\
    N3384 & Surace/3403&  E-S0&  -2.7&  11.59   & 10.78 & 1.272 & 6.56 & 7.07 &61.3$\pm$7.8&44.6$\pm$17.2&\nodata &41.09&40.49&\nodata \\
    N3412 & Fazio/69 & S0 & -2.0 & 11.32  &  10.39 & 1.193 & 6.20 & 6.64& 14.9$\pm$1.9&$\leq18$&$\leq24$ &40.45&$\leq40.07$&$\leq39.84$ \\
    N3489 & Fazio/69 & S0-a & -1.3 & 12.08  &  10.56 & 1.082 & 7.32 & 7.19 &91.6$\pm$6.7&1756$\pm$78&2814$\pm$128 &41.30&42.12&41.96 \\
    N3516 & Rieke/40936 & S0 & -2.0 &  38.37  &  11.11 & 0.649 & \nodata &\nodata& 712$\pm$32&\nodata&\nodata &43.19&\nodata&\nodata \\
    N3522 & Kannappan/30406 & E & -4.8 & 18.79  &   9.75 & \nodata & \nodata &\nodata& 1.7$\pm$0.3&\nodata&\nodata &39.95&\nodata&\nodata \\
    N3557 & Temi/20171&  E&  -4.8 & 45.71 & 11.19 & 1.427 & \nodata  &\nodata& 30.6$\pm$4.5&276$\pm$19&271$\pm$28 &41.98&42.47&42.10 \\
    N3585 & Fazio/69 & E&  -4.7 & 20.04  & 11.27 & 1.357 & \nodata &\nodata& 37.2$\pm$3.8&80.2$\pm$5.2&38.2$\pm$8.2 &41.35&41.22&40.54 \\
    N3593 & Fazio/69 & S0-a & -0.4 & 9.91   &   10.37 & 1.082& 8.30 & 8.98 &1497$\pm$110&1.94e4$\pm$880&\nodata &42.34&42.99&\nodata \\
    N3607 & Fazio/69  &E-S0 & -3.1 & 22.80  & 11.27 & 1.360 & 7.36 & 9.13 &84.9$\pm$6.7&1761$\pm$98&2215$\pm$186 &41.82&42.67&42.41 \\
    N3608 & Fazio/30318 & E & -4.8 & 22.91  &  10.83 & 1.284 & \nodata & 7.51 &18.7$\pm$4.2&69.5$\pm$8.2&103$\pm$21 &41.17&41.27&41.08 \\
    N3610 & Surace/3403 & E & -4.2  & 29.24  & 11.12 & 1.293 & \nodata  &\nodata& 18.4$\pm$4.5&21.1$\pm$7.3&36.5$\pm$16.5 &41.37&40.97&40.84 \\
    N3640 & Fazio/69 &E&  -4.8&  27.04  &  11.20 & 1.361 & \nodata &\nodata& 19.6$\pm$1.6&$\leq18.3$&$\leq26.8$ &41.33&$\leq40.84$&$\leq40.64$ \\
    N3656 & Young/20780 &Sa & -0.5 & 41.0 & 10.86 & \nodata & \nodata & 9.59 &151$\pm$0.2&3447$\pm$179&3911$\pm$295 &42.58&43.47&43.17 \\
    N3706 & Fazio/30318 &E-S0 & -3.3&  37.21 & 11.33 & 1.398 & \nodata  &\nodata& 16.5$\pm$4.4&$\leq18.6$&$\leq22.3$ &41.53&$\leq41.12$&$\leq40.84$ \\
    N3773 & Kennicutt/40204 & S0 & -2.0 & 15.07  &   9.43 & \nodata & 7.94  &\nodata& 139$\pm$9&1591$\pm$183&2379$\pm$232 &41.67&42.27&42.08 \\
    N3870 & Kannappan/30406 & S0 & -2.0 & 14.03  &  9.31 & 0.177 & 8.22 & 7.94 &92.1$\pm$5.1&\nodata&\nodata &41.43&\nodata&\nodata \\
    N3923 & Temi/20171 & E & -4.6 &   19.14 &  11.31 & 1.442 & \nodata &\nodata& 43.7$\pm$6.7&23.8$\pm$5.4&48.4$\pm$7.8 &41.38&40.65&40.60 \\
    N3941 & Rieke/40936 & S0 & -2.0 &  12.19    &  10.59 & 1.272 & 8.74 & 7.44 &23.3$\pm$3.1&\nodata&\nodata &40.71&\nodata&\nodata \\
    N3945 & Fazio/30318 &S0-a&  -1.2  & 21.37&  11.00 & 1.380 & 8.62 &\nodata& 39.9$\pm$4.1&284$\pm$19&2836$\pm$216 &41.43&41.82&42.46 \\
    N3962 & Kaneda/3619  &E & -4.7  & 23.23   & 11.01 & 1.354 & 8.23  &\nodata& 21.8$\pm$5.6&392$\pm$18&544$\pm$24 &41.24&42.03&41.82 \\
    N4026 & Fazio/30318 &S0 & -1.8 & 13.61  &   10.58 & \nodata & 7.98 & 7.75  &23.7$\pm$3.0&141$\pm$13&106$\pm$23 &40.82&41.13&40.64 \\
    N4073 & Fazio/30318 &E  &-4.1 & 79.43 &  11.75 & \nodata & \nodata &\nodata& 5.3$\pm$3.0&$\leq17.7$&$\leq22.6$ &41.70&$\leq41.76$&$\leq41.50$ \\
    N4117 & Kannappan/30406 & S0 & -2.0 & 16.35  &   9.76 & \nodata & \nodata &\nodata& 31.4$\pm$3.4&\nodata&\nodata &41.10&\nodata&\nodata \\
    N4125 & Kennicutt/159 & E&  -4.8 &  27.79  &  11.49 & 1.377 & \nodata &\nodata& 74.7$\pm$6.9&1105$\pm$98&1735$\pm$187 &41.94&42.64&42.48 \\
    N4138 & Fazio/69 & S0-a & -0.8 & 13.8  &  10.35 & 1.038 & 8.86 & 8.32&167$\pm$16     &2505$\pm$162&6144$\pm$642 &41.68&42.39&42.42 \\
    N4150 & Fazio/69 & $S0^0(r)$&  -2.1 &  13.74 & 10.00 & 1.016 & 7.50 & 7.82 &69.3$\pm$8     &1522$\pm$121&1720$\pm$168 &41.29&42.17&41.86 \\
    N4168 & Fazio/69  &E & -4.8 & 33.73    &  11.03 &1.318 & \nodata  &\nodata& 5.2$\pm$1.3&$\leq15$&$\leq26$ &40.95&$\leq40.94$&$\leq40.82$ \\
    N4203 & Fazio/69  &E-S0 & -2.7 & 15.14  & 10.75 & 1.387 & 9.10 & 7.27 &43.3$\pm$3.5&933$\pm$111&2701$\pm$259 &41.17&42.04&42.14 \\
    N4251 & Fazio/69  &S0&  -1.9 & 19.59  & 10.84 & 1.200 & 7.55 & 7.41 &17.3$\pm$2.4&\nodata&\nodata &41.00&\nodata&\nodata \\
    N4261 & Rieke/40936 & E&  -4.8 & 31.62 & 11.44 & 1.464 & \nodata &\nodata& 51.5$\pm$3.2&127$\pm$11&375$\pm$18 &41.89&41.81&41.92 \\
    N4267 & Cote'/3649 &E-S0&  -2.7 & 15.92   & 10.62 & 1.382 & \nodata &\nodata& 11.7$\pm$0.3&$\leq16.9$&$\leq20.9$ &40.65&$\leq40.34$&$\leq40.07$ \\
    N4278 & Fazio/69 & E & -4.8 & 16.07   &  10.88 & 1.319 & 8.84 & 7.50 &44.7$\pm$12.7&829$\pm$68&1491$\pm$128 &41.24&42.04&41.94 \\
    N4291 & Tripp/30603 & E & -4.8 & 26.18   &  10.82 & \nodata & \nodata &\nodata& \nodata&$\leq18$&\nodata &\nodata&$\leq40.08$&\nodata \\
    N4308 & Kannappan/30406 & E & -4.7 & 11.29 &  9.21 & 1.107 & \nodata  &\nodata& 1.49$\pm$0.3&\nodata&\nodata &39.45&\nodata&\nodata \\
    N4344 & Popescu/3475 & S0 & -1.7 & 17.85 &   9.66 & \nodata & \nodata &\nodata& 39.2$\pm$4.5&773$\pm$62&2228$\pm$226 &41.27&42.10&42.20 \\
    N4350 & Cote'/3649 &S0 & -1.8 & 15.92  &   10.63 & 1.292 & \nodata &\nodata& 29.8$\pm$3.3&474$\pm$32&1168$\pm$112 &41.05&41.79&41.82 \\
    N4352 & Treu/30958 & S0 & -2.0 & 30.96   &   10.38 & 1.349 & \nodata &\nodata& 1.7$\pm$0.3&\nodata&\nodata &40.39&\nodata&\nodata \\
    N4365 & Cote'/3649 & E & -4.8 &  17.06 &  11.16 & 1.411 & \nodata  &\nodata& 22.2$\pm$4.7&67.0$\pm$8.0&58.2$\pm$7.3 &40.98&41.00&40.58 \\
    N4371 & Cote'/3649 &S0-a & -1.3 & 14.32   & 10.57 & 1.549 & \nodata  &\nodata& 13.6$\pm$0.7&24.5$\pm$2.7&14.1$\pm$7.0 &40.62&40.41&39.81 \\
    N4374 & Rieke/82 & E & -4.2  &18.37  &   11.38 & 1.594 & \nodata & 7.15 &66.6$\pm$8.6&617$\pm$41&535$\pm$61 &41.53&42.03&41.61 \\
    N4377 & Treu/30958 & E-S0 & -2.6 & 21.26   &  10.47 & 1.292 & \nodata &\nodata& 3.2$\pm$0.3&\nodata&\nodata &40.33&\nodata&\nodata \\
    N4379 & Treu/30958 & E-S0 & -2.8 & 14.19   &  10.15 & 1.269 & \nodata   &\nodata& 4.3$\pm$0.3&\nodata&\nodata &40.11&\nodata&\nodata \\
    N4382 & Cote'/3649 & S0-a & -1.3 & 18.45  &   11.42 & 1.220 & \nodata & 7.34 &52.4$\pm$6.7&23.6$\pm$6.5&24.7$\pm$9 &41.43&40.61&40.27 \\
    N4386 & Tripp/30603&  S0  &-2.0 & 27.04   &  10.81 & 1.331 & \nodata  & \nodata&  \nodata  &178$\pm$14&\nodata &\nodata&41.82&\nodata \\
    N4406 & Fazio/69 & E&  -4.7  & 17.14   &   11.37 & 1.527 & 7.90 &\nodata& 27.5$\pm$3.2     &64$\pm$8&90$\pm$12 &41.08&40.98&40.77 \\
    N4417 & Treu/30958&  S0 & -1.9 & 15.92  &   10.48 & 1.183 & \nodata  &\nodata& 8.8$\pm$1.2&\nodata&\nodata &40.52&\nodata&\nodata \\
    N4421 & Fazio/69 & S0-a  &-0.2   &23.12  &  10.56 & 1.116 & \nodata  &\nodata& 3.3$\pm$0.4    &$\leq25$&$\leq30$ &40.42&$\leq40.84$&$\leq40.56$ \\
    N4434 & Treu/30958  &E  &-4.8  &26.79  &   10.52 & 1.262 & \nodata &\nodata& 4.8$\pm$0.9&\nodata&\nodata &40.71&\nodata&\nodata \\
    N4435 & Cote'/3649 &S0&  -2.1 & 15.92   &  10.83 & 1.508 & \nodata &\nodata& 111$\pm$10.2&2210$\pm$186&3022$\pm$327 &41.62&42.46&42.23 \\
    N4442 & Fazio/69  &S0&  -2.0  & 8.68      & 10.31 & 1.392 & \nodata   &\nodata& 20.2$\pm$2.2&\nodata&\nodata &40.36&\nodata&\nodata \\
    N4458 & Fazio/30318 & E&  -4.8 &  17.22 &  10.10 &1.152 & \nodata   & 7.28&3.3$\pm$1.1&$\leq18.5$&$\leq23.8$ &40.16&$\leq40.45$&$\leq40.20$ \\
    N4459 & Cote'/3649  &S0-a & -1.4 &  16.14   &  10.90 & 1.481 & \nodata & 8.22 &107$\pm$8&2400$\pm$175&3461$\pm$431 &41.62&42.51&42.31 \\
    N4460 & Fazio/69 & S0-a & -0.9  & 9.59   &  9.69 & \nodata & 8.72& 7.27 &274$\pm$31     &4413$\pm$389&4488$\pm$461 &41.58&42.32&41.97 \\
    N4464 & Treu/30958 & S0-a & -0.7    &15.92   &  9.92 & 1.341 & \nodata &\nodata& 2.0$\pm$0.3&\nodata&\nodata &39.88&\nodata&\nodata \\
    N4472 & Fazio/69&  E&  -4.8  &  17.06  &  11.65 & 1.461 & \nodata & 7.10 &74.7$\pm$8.6&61.1$\pm$7.6&66.4$\pm$7.8 &41.51&40.96&40.64 \\
    N4473 & Cote'/3649  &E & -4.7  & 15.71  &  10.88 & 1.504 & \nodata & 7.01 &26.3$\pm$6.7&$\leq18.0$&$\leq25.5$ &40.99&$\leq40.36$&$\leq40.15$ \\
    N4474 & Treu/30958 & S0 & -2.0 & 15.92  &    10.27 & 1.246 & \nodata  &\nodata& 5.4$\pm$0.8&\nodata&\nodata &40.31&\nodata&\nodata \\
    N4476 & Young/20780 & E-S0  &-3.0  &  17.22    & 10.03 & 1.159 & \nodata& 7.98  &35.7$\pm$4.1&528$\pm$89&1323$\pm$135 &41.20&41.90&41.94 \\
    N4477 & Rieke/40936&  S0 & -1.9 &   17.06   & 10.87 & 1.649 & \nodata & 7.55 &12.9$\pm$3.2&$\leq24$&$\leq31$ &40.75&$\leq40.55$&$\leq40.31$ \\
    N4478 & Fazio/30318 &E & -4.8  &  18.11  &   10.52 & 1.477 & \nodata  &\nodata& 12.6$\pm$3.9&$\leq17.9$&$\leq23.7$ &40.79&$\leq40.48$&$\leq40.24$ \\
    N4479 & Treu/30958  &S0&  -1.8&   15.92  &    9.84 & 1.386 & \nodata  &\nodata& 1.2$\pm$0.3&\nodata&\nodata &39.66&\nodata&\nodata \\
    N4482 & Treu/30958  &E&  -4.8&   27.83   &  10.00 & \nodata & \nodata  &\nodata& 0.3$\pm$0.16&\nodata&\nodata &39.54&\nodata&\nodata \\
    N4483 & Treu/30958 & S0-a & -1.3 &  13.69  &    9.91 & 1.217 & \nodata &\nodata& 3.0$\pm$0.3&\nodata&\nodata &39.92&\nodata&\nodata \\
    N4486 & Cote'/3649  &E & -4.3&   16.07    &  11.44 & 1.648 & \nodata  & 7.03 &154$\pm$9&483$\pm$44&896$\pm$91 &41.77&41.81&41.71 \\
    N4489 & Treu/30958  &E & -4.8 &  17.86  &  10.10 & 1.053 & \nodata  &\nodata& 3.2$\pm$0.3&\nodata&\nodata &40.18&\nodata&\nodata \\
    N4494 & Fazio/69  &E & -4.8 &  17.06     & 11.01 & 1.284 & \nodata & 7.24 &34.5$\pm$4.5&316$\pm$12.5&235$\pm$19.6 &41.18&41.67&41.19 \\
    N4515 & Treu/30958&  E-S0  &-3.0  & 15.92     &9.80 & \nodata & \nodata &\nodata& 1.9$\pm$0.2&\nodata&\nodata &39.86&\nodata&\nodata \\
    N4526 & Fazio/69 & S0 & -1.9  & 16.90 &   11.21 & 1.354 & \nodata & 8.60 &267$\pm$12&8134$\pm$632&13706$\pm$1267 &42.06&43.08&42.94 \\
    N4528 & Treu/30958 & S0  &-2.0 &  20.76 &    10.39 & 1.346 & \nodata &\nodata& 4.4$\pm$0.7&\nodata&\nodata &40.45&\nodata&\nodata \\
    N4552 & Kennicutt/159  &E&  -4.6  &  15.34  &  11.03 & 1.625 & \nodata & 7.18 &58.5$\pm$7.8&96.3$\pm$10.2&188$\pm$16 &41.31&41.06&41.00 \\
    N4564 & Fabbiano/20371 & E&  -4.8  &  15.00 &    10.53 & 1.484 & \nodata &\nodata& 23.9$\pm$3.7&\nodata&\nodata &40.90&\nodata&\nodata \\
    N4570 & Cote'/3649 & S0 & -2.0  &  25.90  &  11.10 & 1.253 & \nodata & 7.75 &18.7$\pm$6.0&$\leq23.3$&$\leq29.6$ &41.27&$\leq40.90$&$\leq40.65$ \\
    N4578 & Fazio/69 & S0 & -2.0 &  18.54 &     10.52 & 1.123 & \nodata   &\nodata& 5.2$\pm$0.9&$\leq19$&$\leq26$ &40.43&$\leq40.52$&$\leq40.30$ \\
    N4589 & Kaneda/3619 & E & -4.8  &  21.98 &   10.93 & \nodata & \nodata &\nodata& 15.3$\pm$4.5&275$\pm$16&396$\pm$21 &41.04&41.83&41.63 \\
    N4612 & Fazio/69 & S0 & -2.0 &  26.61  &    10.77 & \nodata & \nodata   &\nodata& 6.0$\pm$0.8&\nodata&\nodata &40.80&\nodata&\nodata \\
    N4621 & Cote'/3649  &E & -4.8  & 18.28   &   11.17 & 1.354 & \nodata  &\nodata& 34.9$\pm$6.3&33.7$\pm$5.7&47.4$\pm$6.6 &41.24&40.76&40.55 \\
    N4623 & Treu/30958 & S0-a & -1.5  & 26.62 &   10.41 & \nodata & \nodata &\nodata& 1.8$\pm$0.2&\nodata&\nodata &40.28&\nodata&\nodata \\
    N4636 & Fazio/69 & E & -4.8  &  17.06   &   11.24 & 1.319 & \nodata & 6.85  &31.8$\pm$5.6&197$\pm$12&185$\pm$24 &41.14&41.47&41.08 \\
    N4638 & Fazio/69 & E-S0 & -2.6 &  21.68  &    10.74 & 1.290 & \nodata &\nodata& 13.9$\pm$2.8&\nodata&\nodata &40.99&\nodata&\nodata \\
    N4649 & Fazio/69 & E&  -4.6 &  17.06  &      11.52 & \nodata & \nodata & 7.78 &108$\pm$10&48.6$\pm$6.8&$\leq30.4$ &41.67&40.86&$\leq40.30$ \\
    N4660 & Cote'/3649 & E & -4.7   &12.82   &   10.28 & 1.268 & \nodata &\nodata& 15.5$\pm$4.3&38.3$\pm$6.2&59.1$\pm$8.2 &40.58&40.51&40.34 \\
    N4694 & Kenney/30945 & S0 & -2.0 & 18.17   &  10.28 & 0.737 & \nodata &\nodata& 110$\pm$9&1545$\pm$112&3278$\pm$334 &41.73&42.42&42.38 \\
    N4696 & Sparks/3506  &E & -3.7  &  39.65  &    11.69 & \nodata & \nodata &\nodata& 23.4$\pm$4.7&133$\pm$13&295$\pm$21 &41.74&42.03&42.02 \\
    N4697 & Surace/3403 & E & -4.8  &  16.22 &    11.22 & 1.237 & \nodata & 7.16 &44.7$\pm$6.7&618$\pm$55&830$\pm$68 &41.24&41.92&41.69 \\
    N4709 & Fazio/30318 &E & -4.4 &  35.32 &    11.24 & 1.397 & \nodata    &\nodata& 8.2$\pm$0.2&$\leq15.5$&$\leq27.0$ &41.18&$\leq41.00$&$\leq40.88$ \\
    N4754 & Cote'/3649 &S0 & -2.5  & 16.83 & 10.83  & 1.275 & \nodata &\nodata& 17.1$\pm$2.9&$\leq12.2$&$\leq22.2$ &40.86&$\leq40.25$&$\leq40.15$ \\
    N4762 & Cote'/3649 &S0  &-1.8 &  15.92   &   10.83 & 1.141 & \nodata &\nodata& 39.1$\pm$6.5&12.1$\pm$2.6&$\leq24.2$ &41.17&40.20&$\leq40.14$ \\
    N4786 & Fazio/30318 &E & -4.2 &  66.16  &  11.50 & \nodata & \nodata &\nodata& 25.3$\pm$1.2&303$\pm$26.0&161$\pm$19 &42.22&42.83&42.20 \\
    N4915 & Surace/3403 & E&  -4.5 &   46.98 &  11.18 & 1.251 & \nodata   &\nodata& 11.7$\pm$5.5&30.6$\pm$4.8&59.7$\pm$8.9 &41.59&41.54&41.47 \\
    N4936 & Fazio/30318 &E & -4.6  & 41.07  & 11.48 & \nodata & \nodata &\nodata& 21.7$\pm$0.9&465$\pm$45.0&894$\pm$76.0 &41.74&42.60&42.53 \\
    N5018 & Surace/3403 & E&  -4.6 &   32.36 &  11.27 & 1.207 & 8.67 &\nodata& 61.7$\pm$9.1&1174$\pm$134&1855$\pm$241 &41.98&42.80&42.64 \\
    N5044 & Temi/20171  &E & -4.7  &  32.36 & 11.28 & 1.389 & \nodata &\nodata& 23.5$\pm$5.3&241$\pm$23&266$\pm$28 &41.57&42.11&41.80 \\
    N5061 & Fazio/30318 &E&  -4.3&   18.28 &   10.96 & 1.209 & 8.04  &\nodata& 32.8$\pm$0.5&18.2$\pm$5.2&$\leq23.5$ &41.21&40.49&$\leq40.25$ \\
    N5077 & Fazio/69  &E & -4.8   & 32.36  &  11.08 & 1.488 & \nodata    &\nodata& 33.2$\pm$5.7&133$\pm$11&155$\pm$18 &41.72&41.85&41.56 \\
    N5173 & Kannappan/30406 & E & -4.7 &  34.99 &  10.42 & \nodata & 9.13 &\nodata& 21.1$\pm$1.5&\nodata&\nodata &41.59&\nodata&\nodata \\
    N5273 & Fazio/69 & S0&  -1.9 &  16.52  &   10.32 & 1.132 & \nodata   &\nodata& 83.1$\pm$4.5&657$\pm$44&836$\pm$79 &41.53&41.96&41.71 \\
    N5322 & Fazio/69 & E & -4.8  &  29.79  &    11.43 & 1.345 & \nodata   &\nodata& 42.2$\pm$6.4&477$\pm$23&687$\pm$77 &41.75&42.34&42.14 \\
    N5338 & Kannappan/30406 & S0  &-2.0&  12.82 &   9.41 & \nodata & 7.34 &\nodata& 35.2$\pm$5.0&\nodata&\nodata &40.94&\nodata&\nodata \\
    N5353 & Fazio/69&  S0 & -2.1  &34.67   &    11.38 & 1.480 & 9.70  &\nodata& 24.7$\pm$3.3&492$\pm$38&490$\pm$52 &41.65&42.48&42.12 \\
    N5419 & Fazio/30318 &E & -4.4 &  53.44   &  11.80 & \nodata & \nodata  &\nodata& 12.9$\pm$0.14&$\leq18.3$&$\leq21.7$ &41.74&$\leq41.43$&$\leq41.14$ \\
    N5481 & Zezas/20140&  E&  -4.0&  30.75  &   10.57 & \nodata & \nodata &\nodata& 1.8$\pm$0.3&10.2$\pm$3.4&$\leq22$ &40.41&40.69&$\leq40.67$ \\
    N5557 & Surace/3403 & E & -4.8   &49.02  &   11.50 & 1.386 & \nodata  &\nodata& 13.4$\pm$5.0&26.4$\pm$7.5&$\leq46$ &41.68&41.51&$\leq41.39$ \\
    N5576 & Surace/3403 &E  &-4.8 & 25.47 &   11.03 & 1.224 & \nodata &\nodata& 16.5$\pm$4.2&$\leq12.7$&$\leq19.8$ &41.20&$\leq40.63$&$\leq40.46$ \\
    N5596 & Kannappan/30406&  S0&  -1.8 &  47.80   &    10.54& 1.219 & \nodata &\nodata& 6.5$\pm$0.5&\nodata&\nodata &41.35&\nodata&\nodata \\
    N5666 & Young/20780 &Sc & 6.4 &   33.45   &  10.33 & \nodata & 9.09  & 8.72 &1503$\pm$0.5&2566$\pm$155&2637$\pm$244 &43.40&43.17&42.82 \\
    N5813 & Fazio/69  &E&  -4.8 &  32.21&     11.40 & 1.418 & \nodata  & 7.64 &15.3$\pm$4.9&61.3$\pm$7.6&38.3$\pm$5.9 &41.37&41.51&40.95 \\
    N5831 & Surace/3403 & E & -4.8 &  27.16   &   10.84 & 1.427 & \nodata & 7.79 &14.8$\pm$4.6&$\leq18.8$&$\leq24.6$ &41.21&$\leq40.85$&$\leq40.61$ \\
    N5845 & Fabbiano/20371  &E&  -4.8&   25.94   &  10.53 & 1.674 & \nodata & 7.45 &7.8$\pm$3.2&108$\pm$12&169$\pm$23 &40.89&41.57&41.41 \\
    N5846 & Fazio/69  &E&  -4.7 &  24.89   &     11.36 & 1.370 & \nodata  & 7.72 &39.3$\pm$6.2&107$\pm$10&129$\pm$12 &41.56&41.53&41.25 \\
    N5866 & Kennicutt/159 &S0-a & -1.2  & 15.35  &   10.97 & 1.108 & 8.08 & 8.64&194$\pm$7.1&8753$\pm$605&17537$\pm$1297.0 &41.83&43.02&42.97 \\
    N5982 & Surace/3403 & E & -4.8 &  40.18  &   11.30 & 1.354 &7.42 & 7.53&13.8$\pm$5.1&39.0$\pm$7.2&65.8$\pm$11.7 &41.52&41.51&41.38 \\
    N6482 & Fazio/69  &E & -4.8 & 54.69 &     2.82 & 1.054 &\nodata &\nodata& 9.5$\pm$0.8    &33.6$\pm$12.1&$\leq45.2$ &41.63&41.71&$\leq41.48$ \\
    N6684 & Fisher/30496 & S0 & -1.8 & 13.93  &    10.82 & 1.189 & \nodata &\nodata& 17.7$\pm$2.0&\nodata&\nodata &40.71&\nodata&\nodata \\
    N6703 & Fazio/69&  E-S0  &-2.8  & 32.06    &  11.06 & 1.360 & \nodata  &\nodata& 20.3$\pm$4.5&43.5$\pm$6.4&21.7$\pm$5.1 &41.49&41.36&40.70 \\
    N6776 & Fazio/30318 &E & -4.1 & 70.41   &  11.43 & 1.293 & \nodata  &\nodata& 16.4$\pm$1.3&126$\pm$12&58.1$\pm$19.3 &42.08&42.50&41.81 \\
    N6849 & Fazio/30318 &E-S0&  -3.3  &84.41   &  11.44 & \nodata & \nodata &\nodata& 3.3$\pm$0.1&$\leq12.6$&$\leq19.8$ &41.55&$\leq41.66$&$\leq41.50$ \\
    N7077 & Kannappan/30406 & E  &-4.1  &17.06   &    9.26 & \nodata & 8.13  &\nodata& 28.7$\pm$3.1&411$\pm$37&452$\pm$42 &41.10&41.79&41.47 \\
    N7176 & Johnson/3596 & E & -4.6 & 32.39  &   11.16 & \nodata & \nodata &\nodata& 2.6$\pm$0.4&\nodata&\nodata &40.61&\nodata&\nodata \\
    N7360 & Kannappan/30406 & E & -5.0 &  67.30 &  10.76 & \nodata & \nodata &\nodata& 9.6$\pm$0.6&\nodata&\nodata &41.81&\nodata&\nodata \\
    N7457 & Fazio/30318 & E-S0 & -2.6 & 13.24    &  10.32 & 1.113& 6.25 & 6.66 &8.4$\pm$2.5&$\leq25.9$&$\leq33.5$ &40.34&$\leq40.37$&$\leq40.12$ \\
    N7619 & Temi/20171 & E & -4.7 &  52.97  &   11.58 & 1.430 & \nodata  &\nodata& 10.2$\pm$4.7&$\leq13.7$&$\leq23.5$ &41.63&$\leq41.29$&$\leq41.17$ \\
    N7626 & Fazio/30318 &E & -4.8 &  39.99 &   11.34 & 1.459 & \nodata   &\nodata& 12.4$\pm$3.6&18$\pm$0.3&$\leq25.3$ &41.47&41.17&$\leq40.96$ \\
    N7785 & Fazio/30318  &E  &-4.8   &55.46  & 11.46 & 1.449 & \nodata   &\nodata& 8.2$\pm$0.7&18$\pm$0.9&$\leq19.0$ &41.58&41.45&$\leq41.12$ \\
        \\
     I0798 & Treu/30958 & E  &-4.4  &   7.66&     8.46 & \nodata & \nodata &\nodata& 0.6$\pm$0.2&\nodata & \nodata &38.72&\nodata & \nodata \\
     I1144 & Kannappan/30406&  E-S0&  -3.0 &     176.20  &   11.57 & \nodata & \nodata  &\nodata& 1.4$\pm$0.3&\nodata & \nodata &41.81&\nodata & \nodata \\
     I1459 & Latter/1712  &E & -4.7    &   29.24   &   11.56 & 1.449& \nodata  &\nodata& 60.9$\pm$4.3&542$\pm$32&627$\pm$72 &41.89&42.38&42.08 \\
     I1639 & Kannappan/30406 & E-S0 & -3.0 &    75.86   &   10.86 & \nodata & \nodata    &\nodata& 3.7$\pm$0.4&\nodata & \nodata &41.50&\nodata & \nodata \\
     I3032 & Treu/30958 & E? & -2.4 &  18.26  &   9.00 & \nodata & \nodata   &\nodata& $\leq0.1$&\nodata & \nodata &$\leq38.70$&\nodata & \nodata \\
     I3101 & Treu/30958&  E? & -2.5  & 31.67   &  9.18 & \nodata & \nodata   &\nodata& $\leq0.1$&\nodata & \nodata &$\leq39.18$&\nodata & \nodata \\
     I3328 & Treu/30958  &E-S0  &-2.9  & 15.71&     9.22 & \nodata & \nodata &\nodata& 0.45$\pm$0.08&\nodata & \nodata &39.22&\nodata & \nodata \\
     I3370 & Kaneda/3619  &E & -4.7  & 26.79  &   11.06 & 1.186& \nodata &\nodata& 32.8$\pm$3.4&680$\pm$51&958$\pm$128 &41.55&42.40&42.19 \\
     I3381 & Treu/30958  &E&  -4.1 &  10.85   &   9.00 & \nodata & \nodata    &\nodata& $\leq0.1$&\nodata & \nodata &$\leq38.24$&\nodata & \nodata \\
     I3383 & Treu/30958 & E?  &-2.7&   27.05    &  9.08 & \nodata & \nodata   &\nodata& $\leq0.1$&\nodata & \nodata &$\leq39.04$&\nodata & \nodata \\
     I3461 & Treu/30958  &E&  -4.1 &  15.60 &    8.77 & 0.30 & \nodata        &\nodata& $\leq0.1$&\nodata & \nodata &$\leq38.56$&\nodata & \nodata \\
     I3468 & Treu/30958  &E & -4.8  & 19.62   & 9.73 & \nodata & \nodata     &\nodata& 0.16$\pm$0.08&\nodata & \nodata &38.96&\nodata & \nodata \\
     I3470 & Treu/30958  &E & -4.0  & 22.71   & 9.59 & \nodata & \nodata    &\nodata& $\leq0.1$&\nodata & \nodata &$\leq38.89$&\nodata & \nodata \\
     I3487 & Treu/30958 & E-S0 & -2.9  & 16.60 &   8.90 & \nodata & \nodata &\nodata& $\leq0.1$&\nodata & \nodata &$\leq38.61$&\nodata & \nodata \\
     I3501 & Treu/30958  &E  &-4.2  & 24.76    &  9.64 & \nodata & \nodata    &\nodata& 0.24$\pm$0.08&\nodata & \nodata &39.34&\nodata & \nodata \\
     I3586 & Treu/30958 & S0-a&  -0.8 &26.06  &  9.33 & \nodata & \nodata   &\nodata& $\leq1.0$&\nodata & \nodata &$\leq40.01$&\nodata & \nodata\\
     I3602 & Treu/30958  &E & -5.0  & 31.43   & 9.61 & \nodata & \nodata     &\nodata& 0.69$\pm$0.11&\nodata & \nodata &40.01&\nodata & \nodata \\
     I3633 & Treu/30958  &E & -3.6   &29.87   &  8.90 & \nodata & \nodata     &\nodata& $\leq0.1$&\nodata & \nodata &$\leq39.12$&\nodata & \nodata \\
     I3652 & Treu/30958  &E & -4.8   &10.32    & 8.96 & \nodata & \nodata    &\nodata& 0.22$\pm$0.08&\nodata & \nodata &38.54&\nodata & \nodata \\
     I3653 & Treu/30958 & S0 & -2.2  & 14.72   & 9.45 & \nodata & \nodata   &\nodata& 0.77$\pm$0.12&\nodata & \nodata &39.40&\nodata & \nodata \\
     I3735 & Treu/30958 & E & -3.5  & 28.08   &9.69 & \nodata & \nodata    &\nodata& $\leq0.1$&\nodata & \nodata &$\leq39.07$&\nodata & \nodata \\
     I3773 & Treu/30958 & E & -4.8  & 16.71   & 9.42 & \nodata & \nodata    &\nodata& 0.34$\pm$0.08&\nodata & \nodata &39.15&\nodata & \nodata \\
     I3779 & Treu/30958 & E & -5.0  & 17.92    & 8.73 & \nodata & \nodata    &\nodata& $\leq0.1$&\nodata & \nodata &$\leq38.68$&\nodata & \nodata \\
     I4296  & Fabbiano/20371 & E & -4.8 &   47.56  &    11.70 & 1.426 & \nodata   &\nodata& 21.4$\pm$8&118$\pm$12&71$\pm$12 &41.86&42.14&41.56 \\
     I4329 & Fazio/30318&E & -3.8 & 61.97  &   10.99 & \nodata & \nodata &\nodata& 7.2$\pm$1.5&$\leq18.5$&$\leq25.8$ &41.62&$\leq41.56$&$\leq41.35$ \\
     I5063  & Werner/86  &S0-a & -0.9   & 45.29   &  11.16 & 1.268 & 9.67 &\nodata& 2170$\pm$78&4425$\pm$321&\nodata &43.82&43.67&\nodata \\
        \\
     E103-35 & Werner/86 & S0-a & -0.3  & 54.20 &  10.68 & \nodata & \nodata      & \nodata& 1693$\pm$81&1774$\pm$112& \nodata &43.87&43.43&\nodata \\
     E428-14 & Werner/86 & S0 & -1.7  & 21.27   &   10.50 & \nodata & \nodata       &\nodata& 1448$\pm$80&4714$\pm$389&\nodata &42.99&43.04&\nodata \\
     E462-15 & Fazio/30318 & E & -4.8  &  84.41  &  11.68 & 1.264 & \nodata &\nodata& 6.1$\pm$0.7&$\leq19.4$&$\leq26.8$ &41.81&$\leq41.85$&$\leq41.63$ \\
     E483-13 & Kennicutt/40204  & E-S0 & -2.8 & 10.05  &  9.02 & \nodata & \nodata   &\nodata& 29.8$\pm$3.8&417$\pm$23&579$\pm$69 &40.65&41.33&41.12 \\
        \\
     U1503   & Young/20780 &E & -4.8 &   74.13& 10.86 & \nodata & 9.63  & 9.32 &39.6$\pm$1.8&362$\pm$66&1384$\pm$139 &42.51&43.01&43.23 \\
     U6570   & Kannappan/30406 & S0-a & -0.2  & 25.35  & 9.86 & \nodata & \nodata   &\nodata& 266$\pm$33&\nodata & \nodata &42.41&\nodata & \nodata \\
     U6637   & Kannappan/30406 & E-S0 & -3.4 & 27.76   &  9.54 & \nodata & \nodata     &\nodata& 18.$\pm$2.3&\nodata & \nodata &41.32&\nodata & \nodata \\
     U6655   & Kannappan/30406 & S0-a & -0.3 & 11.86   &  8.67 & \nodata & 7.60 &\nodata& 18.4$\pm$6  &\nodata & \nodata &40.59&\nodata & \nodata \\
     U6805   & Kannappan/30406 & S0-a & -1.2 & 17.49    & 9.25 & \nodata & \nodata    &\nodata& 42.$\pm$3.6&\nodata & \nodata &41.28&\nodata & \nodata \\
     U7020A & Kannappan/30406 & S0-a & -1.5 & 25.21    &  9.75  & 0.303 & \nodata  &\nodata& 207$\pm$12&\nodata & \nodata &42.29&\nodata & \nodata \\
     U7399A & Treu/30958 & S0 & -2.5 & 22.64 &2.06 & \nodata & \nodata &\nodata& $\leq0.1$&\nodata & \nodata &$\leq38.88$&\nodata & \nodata \\
     U7436   & Treu/30958 & S0-a & -1.0  & 15.49  &  9.19 & \nodata & \nodata &\nodata& $\leq1.0$&\nodata & \nodata &$\leq39.55$&\nodata & \nodata \\
     U7580   & Treu/30958 & S0-a & -1.5  & 9.95    &   8.73 & \nodata & \nodata &\nodata& $\leq0.1$&\nodata & \nodata&$\leq38.17$&\nodata & \nodata \\
     U7854   & Treu/30958  & E & -3.5 & 15.88  &  8.83 & \nodata & \nodata &\nodata& $\leq0.1$&\nodata & \nodata &$\leq38.58$&\nodata & \nodata \\
     U8876   & Kannappan/30406 & S0-a & -0.1 &  33.57 &  10.29 & \nodata & \nodata &\nodata& 1.5$\pm$0.3&\nodata & \nodata  &40.40&\nodata & \nodata \\

  \enddata
\vskip0.3cm
{\scriptsize
NOTES: 
The columns are as follows: 
Column (1). -- Galaxy name. N = NGC, I = IC, E = ESO, U = UGC. 
Column (2). -- Principal Investigator (PI) and Program Identification
number (ID\#) of the Spitzer observing programs from which
galaxies have been selected.
Column (3). -- Galaxy morphological type (Hubble type) from the
HyperLeda Database.
Column (4). -- De Vaucouleurs numerical type, T. Its correspondence
with the Hubble type is defined in RC2. Data values are taken from the
HyperLeda Database. 
Column (5). -- Galaxy distance. When available, distances are taken
from Tonry et al. (2001), otherwise they are from
the NASA Extragalactic Database (NED) and corrected for
$H_0= 70{\rm~km~s}^{-1}{\rm~Mpc}^{-1}$.
Column (6). --  Luminosities in the $Ks$ band, $L_{Ks}$, have been derived
  using the flux density $F_{Ks}$ published
by the 2MASS survey, Skrutskie et al. (2006).
Column (7). -- U-V colors have been derived from the U-B and B-V
colors listed in the HyperLeda database. 
Column (8). -- The neutral hydrogen masses, $M_{HI}$ are taken from
the following publications: Morganti et al. (2006); Sage \& Welch (2006);
Di Serego et al. (2007); Roberts et al. (1991); Huchtmeier (1994);
Huchtmeier et al. (1995); Sadler et al. (200); Knapp \& Raimond (1984).
Column (9). -- Molecular gas masses, $M_{H_2}$ are collected from
the following publications: Combes et al. (2007); Young et al. (2009);
Young et al. (2002);
Sage et al. (2007); Welch \& Sage (2003); Roberts et al. (1991).
Column (10-12). -- MIPS flux densities at 24, 70 and 160 $\mu$m. 
Column (13-15). -- Specific luminosities $L_{\lambda}$ at 24, 70
and 160$\mu$m. In this paper $L_{\lambda}$ represents
$\lambda L_{\lambda}$ in erg s$^{-1}$. 
This differs from our previous notation in which $L_{\lambda}$
represented the Spitzer band width multiplied by the specific
luminosity. \hspace{2.7cm}
}

\end{deluxetable}

\clearpage

\end{document}